\documentclass[showpacs,aps,pra,preprint,superscriptaddress,10pt]{revtex4-1}
\usepackage{geometry}
\usepackage{amsfonts}
\usepackage{graphicx}
\usepackage{breakcites}
\usepackage{amsmath}
\usepackage{hyperref}
\usepackage{cleveref}
\usepackage{mathtools}
\usepackage{physics}
\usepackage{color}
\begin{document}
\title{Feynman path-integral treatment of the Bose polaron beyond the Fr\"{o}hlich model.}

\author{T. Ichmoukhamedov}
\email{timour.ichmoukhamedov@uantwerpen.be}
\affiliation{TQC, Universiteit Antwerpen, Universiteitsplein 1, 2610 Antwerpen, Belgium}
\author{J. Tempere}
\affiliation{TQC, Universiteit Antwerpen, Universiteitsplein 1, 2610 Antwerpen, Belgium}
\affiliation{Lyman Laboratory of Physics, Harvard University, Cambridge, Massachusetts 02138, USA}
\date{\today}

\begin{abstract}
An impurity immersed in a Bose-Einstein condensate is no longer accurately described by the Fr\"{o}hlich Hamiltonian as the coupling between the impurity and the boson bath gets stronger. We study the dominant effects of the two-phonon terms beyond the Fr\"{o}hlich model on the ground-state properties of the polaron using Feynman's variational path-integral approach. The previously reported discrepancy in the effective mass between the renormalization group approach and this theory is shown to be absent in the beyond-Fr\"{o}hlich model on the positive side of the Feshbach resonance. Self-trapping, characterized by a sharp and dramatic increase of the effective mass, is no longer observed for the repulsive polaron once the two-phonon interactions are included. For the attractive polaron we find a divergence of the ground-state energy and effective mass at weaker couplings than previously observed within the Fr\"{o}hlich model.
\end{abstract}


\maketitle

\section{INTRODUCTION}
\label{introduction}

The generic problem of an impurity interacting with a bath of bosonic excitations has been studied for nearly a century. The concept was first introduced by Landau to describe an electron interacting with an ionic lattice in a solid \cite{landau1933electron}. Here, the electron induces a polarization cloud in the lattice with which it combines to form a quasiparticle called a polaron. After much progress in the description of the polaron in various coupling regimes \cite{pekar1946,landau1948effective,Frohlich,LLPoriginal} Feynman proposed a variational path-integral description that interpolated previous weak and strong coupling results \cite{Feynman1955}. Feynman's all-coupling description starts from the Fr\"{o}hlich Hamiltonian \cite{Frohlich} and only incorporates processes of phonon emission or absorption by the electron. In many crystals this is a good approximation and Feynman's method is considered a very successful description of polarons in solids, although for anharmonic phonons, additional processes need to be considered as well \cite{Kussow}.

Polaronic effects are not limited to polar crystals but have also been observed in ultracold atomic gases where an impurity immersed in the quantum gas becomes dressed by the excitations of the gas. 
Quantum gases in general have a large experimental tunability, certainly compared to solids, and provide an ideal ground for the study of polaronic physics throughout various interaction regimes \cite{Tempere2009Feynman}. 
In experiments impurities can be generated by transferring a small fraction of the gas atoms to a different hyperfine state. The impurity-gas interaction can be  tuned from attractive interactions across unitarity towards effective repulsive interactions with Feshbach resonances. The first observations \cite{FermiPolaronExp1,FermiPolaronExp2,Kohstall2012,ZhangFermiPolaron} focused on the Fermi polaron, a single fermionic atom in a sea of opposite-spin fermions. Impurities immersed in a Bose-Einstein condensate (BEC) have also been the subject of a number of experimental studies \cite{Heinze,Hohmann2015,Rentrop2016} and recently the energy of Bose polarons has been measured in two experiments at lower temperatures \cite{Hu,Jorgensen} and near criticality \cite{ZwierleinExperiment2019}. 

The last decade has seen a significant amount of theoretical work towards understanding Bose polarons. At weaker impurity-boson coupling strengths, when the density of excitations in the gas is small, the Hamiltonian describing the problem reduces to the Fr\"{o}hlich Hamiltonian \cite{Tempere2009Feynman,CasteelsFrohlichMPLLP,CasteelsPekar,
Casteels2013,Blinova2013,Dasenbrook2013,Vlietinck2015,
DemlerRGFrohlich,DemlerGaussianFrohlich,AllCouplingRGFrohlich,Ardila_2018}. However, it has been pointed out using \textit{T}-matrix calculations \cite{SPRath} and direct perturbative calculations \cite{ChristensenPert} that processes beyond those included in the Fr\"{o}hlich model cannot be neglected at stronger coupling. In particular two-phonon processes, that we will refer to as ``extended Fr\"{o}hlich interactions'' come into play. These are characterized in terms of Feynman diagrams by a vertex joining two impurity lines and two phonon lines. Such processes have also been shown to be crucial for the description of Rydberg polarons \cite{RydbergSchmidt}. This realization has lead to intense theoretical efforts to explicitly incorporate the extended Fr\"{o}hlich interactions in various analytical methods that were developed for and applied to the Fr\"{o}hlich model \cite{BruunEfimov,SchadilovaDynamics,DemlerRG2017,VanLoon2018}. Quantum Monte Carlo (QMC) methods have also been utilized on the full impurity-boson Hamiltonian for the single polaron \cite{ArdilaMC2015,ArdilaMC2019} and for the bipolaron \cite{BipolaronMC} as well. In several of these studies, the excitations of the Bose condensate are described within the framework of the Bogoliubov approximation and it remains an open question as to the range of validity of this approximation.

Feynman's path-integral method, the most successful approach for the solid state polaron, has thus far not been extended to incorporate the two-phonon processes beyond the Fr\"{o}hlich model. Moreover, while it is known to give extremely accurate results for the polaronic ground-state properties in crystals, its accuracy for the Bose polaron has been questioned as it consistently displays quantitative discrepancies from diagrammatic Monte Carlo (MC) calculations within the Fr\"{o}hlich-Hamiltonian model at high-momentum cutoff \cite{Vlietinck2015}. This concern is largely addressed in \cite{DemlerRGFrohlich}, where a renormalization group (RG) approach is used to point out that logarithmic divergences that have not been regularized in the MC results in \cite{Vlietinck2015} are not being captured within Feynman's approach nor within the mean-field (MF) approach, explaining the large discrepancies. Note that beyond weak coupling Feynman's method still provides a much lower bound for the polaronic contribution to the energy than the mean-field approach. The absence of this logarithmic divergence however suggests that Feynman's method does not fully capture the same quantum fluctuations as the ones studied in the RG theory at high momentum. At smaller cutoff scales, where the logarithmic regularization is presumably of less importance, Feynman's approach still yields an energy very close to the MC for the Bose polaron. The other concern pointed out in \cite{DemlerRGFrohlich,GrusdtReview,2DRGtheory,AllCouplingRGFrohlich} is that within the Fr\"{o}hlich model, Feynman's method predicts a very sudden and sharp transition to the strong coupling regime, in particular for the effective mass, while this is not expected in the mean-field, RG or other variational models \cite{DemlerGaussianFrohlich}. In this paper we will show that for the repulsive polaron this artifact is no longer present in Feynman's approach once extended Fr\"{o}hlich interactions are added and the transition is smoothed out.

The main aim of this work is to improve the Feynman path-integral method for Bose polarons \cite{Tempere2009Feynman} so as to take the extended Fr\"{o}hlich interactions into account. In \cref{section1} we start by outlining the general problem in second quantization and find the
corresponding Lagrangian required for the path-integral approach. The extended Fr\"{o}hlich interactions lead to additional quadratic but non-diagonal contributions in the phonon position and velocity variables in the Lagrangian. We perform the path integration over the phonon variables
within the Bogoliubov approximation in \cref{section2} by writing the additional position-dependent extended interaction terms as a full perturbative series with respect to the Fr\"{o}hlich model. Applying the Jensen-Feynman inequality yields an expression for the variational free
energy containing impurity density operator correlations at different times up to an arbitrarily large order. To retrieve an analytic result a random phase-approximation (RPA) of these correlations can be made. In the limit of weak coupling and zero temperature the approximated variational free energy and the effective mass reduce to the extended Fr\"{o}hlich
mean-field results at zero polaron momentum \cite{SchadilovaDynamics,VanLoon2018}. This provides a justification of the RPA at weak to intermediate coupling. We find that the addition of the extended Fr\"{o}hlich interactions allows us to fully regularize the
contact potential in a manner similar as in \cite{SchadilovaDynamics}. In \cref{section3} we compare our results for the repulsive polaron to those obtained within the Fr\"{o}hlich model \cite{Tempere2009Feynman} and find significant differences at strong coupling in the ground-state
energy, effective mass and polaron radius. The effective mass no longer exhibits a sudden sharp increase, but slowly transitions to the strong coupling regime, accompanied by a finite non-zero polaron radius. This provides an indication against the self-trapping of the repulsive polaron that was 
present in the Fr\"{o}hlich model. We also directly compare our results to other recent theoretical models across the resonance. For the energy of the repulsive polaron we find good agreement with the QMC data points presented in \cite{ArdilaMC2019} but quantitative discrepancies with the RG approach \cite{DemlerRG2017}, especially in the limit of strong coupling. Nevertheless, the
effective mass of the repulsive branch is in good agreement with the predictions of the RG approach. 
On the attractive branch we find that at weak coupling the variational landscape contains a local minimum that is identified as the polaronic state, but also separated poles. At a critical coupling strength the local minimum vanishes and the polaron energy and effective mass diverge. Similar behavior is observed within RG but not within QMC and our results provide another indication that this early divergence on the attractive branch could be related to the shortcomings of the Bogoliubov approximation \cite{DemlerRG2017}, or the importance of correctly capturing Efimov physics in this regime \cite{BruunEfimov,SunEfimov,YoshidaEfimov}.

\section{The Hamiltonian and Lagrangian of an impurity in a condensate}
\label{section1}

The full Hamiltonian describing $N_{I}$ impurities with mass $m_I$ immersed in a gas of bosons with mass $m_b$ confined to a box with volume $V$ is given by:
\begin{align}
\hat{H} = \sum_{\mathbf{j}}^{N_I} \frac{ \hat{\mathbf{p}}_j^2}{2m_I} + \sum_{\mathbf{k}} E_{\mathbf{k}} \hat{a}^{\dagger}_{\mathbf{k}} \hat{a}_{\mathbf{k}}+ \frac{1}{2} \sum_{\mathbf{q}} V_{\mathbf{q}}^{BB} \sum_{\mathbf{k},\mathbf{k^{\prime}}} \hat{a}^{\dagger}_{\mathbf{k+q}} \hat{a}^{\dagger}_{\mathbf{k'-q}} \hat{a}_{\mathbf{k'}} \hat{a}_{\mathbf{k}} + \sum_{\mathbf{q}} V_{\mathbf{q}}^{IB} \hat{\rho}_{\mathbf{q}} \sum_{\mathbf{k}} \hat{a}^{\dagger}_{\mathbf{k-q}} \hat{a}_{\mathbf{k}} + \frac{1}{2} \sum_{i \neq j }^{N_I}
U(\mathbf{\hat{r}}_i-\mathbf{\hat{r}}_j).
\label{Ham1}
\end{align}
The bosonic creation and annihilation operators 
$ \hat{a}^{\dagger}_{\mathbf{k}}, \hat{a}_{\mathbf{k}}$ are written in second quantization, while the impurities are considered in first quantization with impurity position and momentum operators 
$\mathbf{\hat{r}}_j, \mathbf{\hat{p}}_j$. The first term describes the kinetic energy of the impurities and the second term describes the kinetic energy of the bosonic atoms with $E_{\mathbf{k}}=(\hbar k)^2/(2m_b)$. The third term represents the interactions between the bosonic atoms, where $V^{BB}_{\mathbf{q}}$ is the Fourier representation of the boson-boson potential. The fourth term represents the impurity-boson interaction and similarly contains the impurity-boson potential $V_{\mathbf{q}}^{IB}$ and the impurity density operator $\rho_{\mathbf{q}}= \sum_{j} \exp \left( i \mathbf{q} \cdot \mathbf{\hat{r}}_j\right) $. The final term is the impurity-impurity interaction potential $U(\mathbf{\hat{r}}_i-\mathbf{\hat{r}}_j)$. In what follows we will consider the interparticle interactions to be contact interactions $V_{\mathbf{q}}^{BB}=g_{bb}/V$ and $V_{\mathbf{q}}^{IB}=g_{ib}/V$. 

Following the Bogoliubov approximation for a weakly interacting 
BEC we assume the $\mathbf{k}=0$ mode to be macroscopically occupied 
with $N_0$ bosons and rewrite the Hamiltonian in terms of the 
Bogoliubov-transformed operators 
$\hat{\alpha}^{\dagger}_{\mathbf{k}},\hat{\alpha}_{\mathbf{k}}$ 
that create resp.~annihilate a Bogoliubov excitation. Keeping up 
to quadratic order in the Bogoliubov operators one obtains 
\cite{BruunEfimov,SchadilovaDynamics,DemlerRG2017,RydbergSchmidt}:
\begin{align}
\hat{H} = & E_{0}+ \frac{g_{ib} N_I N_B}{V}  +\sum_{\mathbf{k }} \epsilon({\mathbf{k}}) \hat{\alpha}^{\dagger}_{\mathbf{k}} \hat{\alpha}_{\mathbf{k}}+ \frac{\sqrt{N_0}g_{ib}}{V} \sum_{\mathbf{k  }}  \hat{\rho}_{\mathbf{k}} V_{\mathbf{k}} \left( \hat{\alpha}^{\dagger}_{\mathbf{-k}} + \hat{\alpha}_{\mathbf{k}} \right)  \nonumber \\
&+\frac{g_{ib}}{V}  \sum_{\mathbf{k,s }} \hat{\rho}_{\mathbf{k-s}} W^{(1)}_{\mathbf{k},\mathbf{s}} \hat{\alpha}^{\dagger}_{\mathbf{s}} \hat{\alpha}_{\mathbf{k}} + \frac{1}{2} \frac{g_{ib}}{V}   \sum_{\mathbf{k,s }}   \hat{\rho}_{\mathbf{k-s}} W^{(2)}_{\mathbf{k},\mathbf{s}} \left( \hat{\alpha}^{\dagger}_{\mathbf{s}}  \hat{\alpha}^{\dagger}_{\mathbf{-k}}  + \hat{\alpha}_{\mathbf{k}} \hat{\alpha}_\mathbf{-s} \right)  \nonumber \\
& +\sum_{\mathbf{j}}^{N_I} \frac{ \hat{\mathbf{p}}_j^2}{2m_I} +  \frac{1}{2} \sum_{i \neq j }^{N_I} U(\mathbf{r_i}-\mathbf{r_j}).
\label{Ham2}
\end{align}
The first term, $E_0$, is the ground-state energy of the BEC, the `vacuum' energy for the Bogoliubov excitations. The second term is the first-order impurity-boson interaction energy where $N_B$ is the total number of bosons. The third term is the kinetic energy of the Bogoliubov excitations with the dispersion relation $\epsilon(\mathbf{k})=\hbar \omega_{\mathbf{k}}=\sqrt{E_{\mathbf{k}}\left(E_{\mathbf{k}}+2 n_0 g_{bb} \right)}$ where $n_0=N_0/V$. The fourth term is the Fr\"{o}hlich interaction term, characterized by an effective interaction potential $V_{\mathbf{k}} = \left[E_{\mathbf{k}}/ \left({E_{\mathbf{k}}+2g_{bb}n_0} \right)\right]^{1/4}$. 
The next two terms 
represent the extended Fr\"{o}hlich interactions where the impurity interacts with two excitations simultaneously, described by effective potentials $W_{\mathbf{k,k'}}^{(1)}= \frac{1}{2} \left( V_{\mathbf{k}}V_{\mathbf{k'}}+ V_{\mathbf{k}}^{-1} V_{\mathbf{k'}}^{-1} \right)$  and $W_{\mathbf{k,k'}}^{(2)}= \frac{1}{2} \left( V_{\mathbf{k}}V_{\mathbf{k'}}- V_{\mathbf{k}}^{-1} V_{\mathbf{k'}}^{-1} \right)$. In the rest of this work we will not consider depletion effects and approximate $N_B$ by the number of condensed atoms $N_0$. 

To apply the path-integral formalism, the Lagrangian corresponding to
(\ref{Ham2}) is needed. For this, the Bogoliubov creation and annihilation operators are combined into effective phonon coordinates $Q_{\mathbf{k}}$ following the standard prescription:
\begin{equation}
Q_{\mathbf{k}}= \sqrt{\frac{\hbar}{2M \omega_\mathbf{k}}}\left(\hat{\alpha}_\mathbf{k}+\hat{\alpha}^{\dagger}_{\mathbf{-k}}\right).
\end{equation}
Here, $M$ is an arbitrary phonon mass which will not appear in the effective action of the polaron, not to be confused with the variational mass of the model system. The Lagrangian, written in terms of these phonon degrees of freedom is derived in appendix (\ref{Appendix A}) for a single impurity and is given by:
\begin{align}
L = & \frac{ m_I \mathbf{\dot{r}}^2}{2} +  \frac{M}{2} \sum_{\mathbf{k}} \dot{Q}_{\mathbf{k}}^{*} \dot{Q}_{\mathbf{k}}- \sum_{\mathbf{k}} \frac{M \omega_{\mathbf{k}}^2}{2} Q_{\mathbf{k}}^* Q_{\mathbf{k}}  -\frac{\sqrt{N_0}g_{ib}}{V} \sum_{\mathbf{k  }}  \rho_{\mathbf{k}}  \sqrt{\frac{2M \omega_{\mathbf{k}}}{\hbar}}V_{\mathbf{k}} Q_{\mathbf{k}}  \nonumber \\
&-\frac{g_{ib}}{V} \frac{M}{2}  \sum_{\mathbf{k,s }}  \rho_{\mathbf{k-s}} V_{\mathbf{k}} V_{\mathbf{s}}  \frac{ \sqrt{ \omega_{\mathbf{k}} \omega_{\mathbf{s}} }}{\hbar} Q_{\mathbf{s}}^* Q_{\mathbf{k}} -\frac{g_{ib}}{V} \frac{M \eta}{2} \sum_{\mathbf{k,s}} \frac{V_{\mathbf{k}}^{-1} V_{\mathbf{s}}^{-1}}{\hbar \sqrt{\omega_{\mathbf{k}} \omega_{\mathbf{s}}}}  \rho_{\mathbf{k}-\mathbf{s}}\dot{Q}_{\mathbf{k}}\dot{Q}_{\mathbf{s}}^*,
\label{Lag1}
\end{align}
where 
\begin{equation}
 \eta=\left(1 +  \frac{g_{ib}}{V} \sum_{\mathbf{k}}
 \frac{V_{\mathbf{k}}^{-2}}{\hbar \omega_{\mathbf{k}}} 
 \right)^{-1},
\end{equation}
and $\mathbf{r}$ is the impurity coordinate. It is clear that if we do not perform any regularization procedures of the 
contact interaction, $\eta$ vanishes at infinite cutoff. The first four 
terms of (\ref{Lag1}) correspond to the Lagrangian of the 
Fr\"{o}hlich model and yield the Fr\"{o}hlich action $\mathcal{S}^{F}$. The additional terms in (\ref{Lag1}) take into account the extended interactions beyond the Fr\"{o}hlich model and consist of a part depending on the phonon coordinates $Q_{\mathbf{k}}$ multiplied by $V_{\mathbf{k}}$, and a part depending on the phonon velocities $\dot{Q}_{\mathbf{k}}$ along with $V_{\mathbf{k}}^{-1}$. Within a mean-field approach where the excitation operators acquire a polaronic shift $\hat{\alpha}_{\mathbf{k}} \rightarrow \hat{\alpha}_{\mathbf{k}} - f_{\mathbf{k}}$, these velocity-dependent terms can be shown to arise due to a non-zero imaginary contribution from $f_{\mathbf{k}}$ and vanish for the saddle-point solution \cite{SchadilovaDynamics,VanLoon2018}, while the terms containing $V_{\mathbf{k}}$ arise due to the real part of $f_{\mathbf{k}}$ and have a non-negligible contribution to the ground-state energy resulting in a resonance shift. In the RG approach \cite{DemlerRG2017} it is pointed out that the RG coupling constant corresponding to the $V_{\mathbf{k}}^{-1}$ terms has a small effect on the polaron wavefunction, but is expected to be important when considering other qualitative properties such as the lifetime of the polaron due to the appearance of bound states at lower energies. These considerations are, however, beyond the scope of this work and we will only consider the position-dependent terms of the extended interactions. In this way we capture the same effects as the mean-field treatment but treat them beyond the mean-field level. Hence, in the remainder of this work, we consider the Euclidean polaron action functional,
\begin{equation}
    S = S^F + \frac{g_{ib}}{V} \frac{M}{2}  \sum_{\mathbf{k,s }}  \int_0^{\hbar \beta} d\tau \rho_{\mathbf{k-s}}(\tau) V_{\mathbf{k}} V_{\mathbf{s}}  \frac{ \sqrt{ \omega_{\mathbf{k}} \omega_{\mathbf{s}} }}{\hbar} Q_{\mathbf{s}}^*(\tau) Q_{\mathbf{k}}(\tau),  
    \label{fullaction}
\end{equation}
where
\begin{equation}
    S^F=\int_{0}^{\hbar \beta} \left(  \frac{ m_I \mathbf{\dot{r}}^2}{2} +  \frac{M}{2} \sum_{\mathbf{k}} \dot{Q}_{\mathbf{k}}^{*} \dot{Q}_{\mathbf{k}}+ \sum_{\mathbf{k}} \frac{M \omega_{\mathbf{k}}^2}{2} Q_{\mathbf{k}}^* Q_{\mathbf{k}}  +\frac{\sqrt{N_0}g_{ib}}{V} \sum_{\mathbf{k }}  \rho_{\mathbf{k}}  \sqrt{\frac{2M \omega_{\mathbf{k}}}{\hbar}}V_{\mathbf{k}} Q_{\mathbf{k}} \right) d\tau
    \label{FrohlichAction}
\end{equation}
is the action of the Frohlich model.

\section{Feynman's variational path-integral approach }
\label{section2}

The free energy $F$ of the polaron can be expressed as a path integral over the impurity and phonon degrees of freedom, weighted by the exponent of the action functional $\mathcal{S}\left[\mathbf{r}, Q_{\mathbf{k}}  \right]$  corresponding to Lagrangian (\ref{Lag1}):
\begin{equation}
 e^{-\beta F}= \int \mathcal{D} \mathbf{r} \int \mathcal{D} \lbrace Q_{\mathbf{k}} \rbrace e^{-\mathcal{S}\left[\mathbf{r}, Q_{\mathbf{k}}  \right] / \hbar} = \int \mathcal{D} \mathbf{r} e^{-\mathcal{S}_{\textrm{eff}} \left[ \mathbf{r} \right] / \hbar} ,
\end{equation}
where
$\beta=(k_B T)^{-1}$ is the inverse temperature and $\mathcal{S}_{\textrm{eff}}\left[ \mathbf{r} \right]$ is the effective imaginary-time action where the phonon degrees of freedom have been integrated out. For a variational model system with action $\mathcal{S}_0$ and free energy $F_0$ the Jensen-Feynman inequality provides a variational upper bound for the free energy of the polaron \cite{Feynman1955} (see \cite{FeynmanStatisticalMechanics,Kleinert} for details):
\begin{equation}
F \leq F_0 + \frac{1}{\hbar \beta} \expval{\mathcal{S}-\mathcal{S}_0}_0,
\label{JensenInequality}
\end{equation}
where $\expval{...}_0$ is the expectation value with respect to the variational model system. 

\subsection{Outline within the Fr\"{o}hlich model}

Within the Fr\"{o}hlich model (so, excluding extended interactions) the effective action for a Bose polaron is given by \cite{Tempere2009Feynman}:
\begin{equation}
\mathcal{S}_{\textrm{eff}}^{F}=  \int_{0}^{\hbar \beta} \frac{m_I \dot{\mathbf{r}}^2}{2} dt - \frac{1}{V} \sum_{\mathbf{k}} \frac{g_{ib}^2 n_0}{2 \hbar } \ V_{\mathbf{k}}^2 \int_{0}^{\hbar \beta}  d\tau \int_{0}^{\hbar \beta}  d\sigma   \mathcal{G}\left(\mathbf{k},|\tau-\sigma| \right) \rho_{\mathbf{k}}(\tau) \rho_{\mathbf{k}}^{*}(\sigma),
\label{effectiveactionFrohlich}
\end{equation}
with $n_0$ the BEC density and
$\rho_{\mathbf{k}}(\tau) = 
  \exp\left[i \mathbf{k} \cdot \mathbf{r}(\tau) \right]$. 
As a consequence of integrating out the phonon variables the effective action (\ref{effectiveactionFrohlich}) now contains a retarded interaction mediated by the Green's function of the Bogoliubov excitations:
\begin{equation}
\mathcal{G}(\mathbf{k},u) = \frac{\cosh \left[ \omega_{\mathbf{k}} \left( |u| - \hbar \beta/2 \right)\right]}{\sinh \left( \omega_{\mathbf{k}} \hbar \beta/2 \right)}.
\end{equation}

We will consider the same variational model system $\mathcal{S}_0$ as in \cite{Feynman1955,Tempere2009Feynman}. 
This system physically corresponds to a particle with the same mass as the impurity $m_I$ coupled to a second mass $M$ by a spring constant $MW^2$, $M$ and $W$ being variational parameters. 
Hence, the variational model action is given by:
\begin{equation}
\mathcal{S}_0 =  \int_{0}^{\hbar \beta} \frac{m_I \dot{\mathbf{r}}^2}{2} dt  + \frac{M W^3}{8} \int_0^{\hbar \beta} d\tau \int_0^{\hbar \beta} d\sigma \frac{\cosh \left[ W \left( |\tau-\sigma|- \hbar \beta/2 \right)\right]}{\sinh \left( W \hbar \beta/2 \right)} \left[ \mathbf{r}(\tau) - \mathbf{r}(\sigma)\right]^2.
\end{equation}
The free energy of the model system $F_0$ can be straightforwardly calculated and the variational upper bound for the polaron energy (\ref{JensenInequality}) can be minimized as a function of $M$ and $W$. For the Fr\"{o}hlich model this has been done by Feynman for a polaron in a crystal \cite{Feynman1955} and more recently applied to a Bose polaron as well \cite{Tempere2009Feynman}.
In the following subsection we will go beyond the Fr\"{o}hlich model for the Bose polaron by including the extended Fr\"{o}hlich interactions in the effective action of the polaron system $\mathcal{S}_{\textrm{eff}}$.

\subsection{Perturbative expansion for the beyond Fr\"ohlich terms}
The effective action corresponding to (\ref{fullaction}) is obtained by integrating out the phonon degrees of freedom. First, $\exp({-\mathcal{S}}/ \hbar)$ is factorized in the
Fr\"ohlich contribution $\exp({-\mathcal{S}^F} / \hbar)$ and the beyond-Fr\"ohlich part:
\begin{equation}
e^{-\mathcal{S}_{\textrm{eff}} / \hbar } = \int \mathcal{D} \lbrace Q_{\mathbf{k}} \rbrace \exp \left( - \frac{g_{ib}}{V} \frac{M}{2}  \sum_{\mathbf{k,s }} V_{\mathbf{k}} V_{\mathbf{s}}  \frac{ \sqrt{ \omega_{\mathbf{k}} \omega_{\mathbf{s}} }}{\hbar^2} \int_0^{\hbar \beta} d\tau  \rho_{\mathbf{k-s}}(\tau)Q_{\mathbf{k}}(\tau)  Q_{\mathbf{s}}^*(\tau)  \right) e^{-\mathcal{S}^{F}/ \hbar}.
\label{pert1}
\end{equation}
The idea is to take into account the exponential of the beyond-Fr\"ohlich terms in (\ref{pert1}) perturbatively through a series expansion of the exponential and a subsequent integration over the phonon degrees of freedom. The terms in the resulting perturbation series can be obtained more straightforwardly with the generating functional formalism. The generating functional is obtained by adding source terms (and a prefactor that will simplify the algebra) to the Fr\"{o}hlich action:
\begin{equation}
   S^{F}[J_{\mathbf{k}}]= S^{F} + \frac{1}{2} \frac{\sqrt{N_0}g_{ib}}{V}   \int_{0}^{\hbar \beta}  \sum_{\mathbf{k}}\sqrt{\frac{2M \omega_{\mathbf{k}}}{\hbar}}V_{\mathbf{k}} \left[  Q_{\mathbf{k}}(\tau) J_{\mathbf{k}}(\tau) + Q^*_{\mathbf{k}}(\tau)J^*_{\mathbf{k}}(\tau) \right] d\tau.
   \label{FrohlichActionSource}
\end{equation}
The source terms resemble the Fr\"ohlich impurity-phonon interaction term, and can be added to it. Hence, including the source terms in the Fr\"{o}hlich action amounts to shifting
$\rho_{\mathbf{k}}$ to $\rho_{\mathbf{k}}+J_{\mathbf{k}}$. The generating functional is then obtained by integrating out the phonon degrees of freedom. The resulting effective action of the Fr\"{o}hlich model including source terms then becomes:
\begin{equation}
S_{\textrm{eff}}^{F} \left[ J_{\mathbf{k}} \right] =  \int_{0}^{\hbar \beta} \frac{m_I \dot{\mathbf{r}}^2}{2} dt -  \sum_{\mathbf{k}} \frac{g_{ib}^2 n_0}{2 \hbar V} \ V_{\mathbf{k}}^2 \int_{0}^{\hbar \beta}  d\tau \int_{0}^{\hbar \beta}  d\sigma   \mathcal{G}\left(\mathbf{k},|\tau-\sigma| \right) \left[\rho_{\mathbf{k}}(\tau) + J_{\mathbf{k}}(\tau) \right] \left[ \rho_{\mathbf{k}}^{*}(\sigma) + J_{\mathbf{k}}^{*}(\sigma)\right].
\label{effectiveactionFrohlichSource}
\end{equation}
In the series expansion of the exponential in (\ref{pert1}) the phonon position 
variables $Q_{\mathbf{k}}$ can be replaced by functional derivatives with respect to 
$J_{\mathbf{k}}$, which can be brought out of the functional integral over $Q_{\mathbf{k}}$. After performing the path integral over the phonon variables one 
is left with the following expression:
\begin{equation}
e^{-\mathcal{S}_{\textrm{eff}} / \hbar}  = \left. \sum_{n=0}^{\infty} \frac{(-1)^n}{n!} \left[ \frac{\hbar }{ g_{ib}n_0}  \sum_{\mathbf{k,s }}    \int_0^{\hbar \beta} d\tau  \rho_{\mathbf{k-s}}(\tau)  \frac{\delta}{\delta J_{\mathbf{k}}(\tau)}  \frac{\delta}{\delta J^*_{\mathbf{s}}(\tau)} \right]^n  e^{-\mathcal{S}_{\textrm{eff}}^{F}\left[  J_{\mathbf{k}} \right] / \hbar} \hspace{3pt} \right|_{J_{\mathbf{k}}=0}.
\label{pert2}
\end{equation}
We will now provide an overview of the structure of the various terms appearing in the generating functional series (\ref{pert2}) and argue which terms can be neglected. They can be classified in three categories.
\subsubsection{Vacuum energy terms}
It is illustrative to consider the $n=1$ order in the expansion. After the first $\frac{\delta}{\delta J^*_{\mathbf{s}}(\tau)}$ in (\ref{pert2}) is applied to the exponential, one obtains
\begin{align}
    e^{-\mathcal{S}_{\textrm{eff}} / \hbar}= \left[ 1-\frac{g_{ib}}{2 \hbar V} \sum_{\mathbf{k,s }} V_{\mathbf{s}}^2  \int_0^{\hbar \beta} d\tau  \rho_{\mathbf{k-s}}(\tau) \frac{\delta}{\delta J_{\mathbf{k}}(\tau)}    \int_0^{\hbar \beta}  \mathcal{G}\left(\mathbf{s},\tau-\sigma \right) \left[ \rho_{\mathbf{s}}(\sigma)+ J_{\mathbf{s}}(\sigma)\right] d\sigma \right] \left. e^{-\mathcal{S}_{\textrm{eff}}^{F}\left[  J_{\mathbf{k}} \right] / \hbar} \hspace{3pt} \right|_{J_{\mathbf{k}}=0}.
    \label{firstorderPert}
\end{align}
Now there is a choice whether to apply the second operator $\frac{\delta}{\delta J_{\mathbf{k}}(\tau)}$ to the exponential again or to the $J_{\mathbf{s}}(t)$ in front. The former option leads to terms which will be discussed in the next subsection. The latter option fully eliminates the impurity variable and results in:
\begin{equation}
    e^{-\mathcal{S}_{\textrm{eff}}  / \hbar}=\left[1- \frac{g_{ib} \beta}{2  V} \sum_{\mathbf{k }}       V_{\mathbf{k}}^2   \mathcal{G}\left(\mathbf{k},0 \right) \right] \left. e^{-\mathcal{S}_{\textrm{eff}} ^{F}\left[  J_{\mathbf{k}} \right] / \hbar} \hspace{3pt} \right|_{J_{\mathbf{k}}=0}.
\label{VacPol1}
\end{equation}
In every order $n$ of the expansion there will be $0 \leq j \leq n$ operator pairs in which each individual pair is applied in the same way as in this example to eliminate the impurity variable and merely yield the term in (\ref{VacPol1}) to the power $j$ multiplied with a combinatorial factor. This allows to separate these terms and perform their complete series summation, given that they are not to be counted from this point on. The summation results in the following contribution to the effective action:
\begin{equation}
\delta S_{\textrm{eff}} = \frac{g_{ib}  \hbar \beta}{2 V} \sum_{\mathbf{k}} V_{\mathbf{k}}^2 \mathcal{G}(\mathbf{k},0). 
\label{VacPol2}
\end{equation}
At zero temperature the corresponding energy shift is given by:
\begin{equation}
    \Delta E_{vac} =  \frac{g_{ib} }{2  V} \sum_{\mathbf{k }}       V_{\mathbf{k}}^2 .
    \label{VacPol3}  
\end{equation}
This contribution is precisely of the same type as the divergent terms arising from non-commuting variables in the derivation of the Lagrangian in appendix (\ref{Appendix A}). First-order corrections in $g_{ib}$ to the ground-state energy are not observed in a rigorous perturbative calculation \cite{ChristensenPert} and we do not expect these terms to be of physical significance. Furthermore note that (\ref{VacPol3}) is UV divergent and can not be regularized by taking the cutoff dependence of $g_{ib}$ into account.
Therefore we will discard contribution (\ref{VacPol2}) in the rest of our calculations.
\subsubsection{Scattering terms}
In the previous example of expansion order $n=1$ we could have also applied the second functional derivative to the exponential again in (\ref{firstorderPert}) to find:
\begin{equation}
     e^{-\mathcal{S}_{\textrm{eff}}  / \hbar}= \left(1-g_{ib} O_1 \left[J_{\mathbf{k}} \right] \right) \left. e^{-\mathcal{S}_{\textrm{eff}} ^{F} \left[J_{\mathbf{k}} \right] / \hbar}\hspace{3pt} \right|_{J_{\mathbf{k}}=0},
\end{equation}
where $O_1 \left[J_{\mathbf{k}} \right]$ is given by
\begin{align}
   O_1 \left[J_{\mathbf{k}} \right]=\frac{g_{ib}^2 n_0}{\hbar \left(2 \hbar V \right)^2} &\sum_{\mathbf{k_1,k_2 }} V_{\mathbf{k_1}}^2 V_{\mathbf{k_2}}^2  \int_0^{\hbar \beta} d\tau_1 \int_0^{\hbar \beta} d\tau_2 \int_0^{\hbar \beta} d\tau_3 \left[ \rho^{*}_{\mathbf{k_1}}(\tau_1) +J^{*}_{\mathbf{k_1}}(\tau_1)\right]  \nonumber \\
    &    \rho_{\mathbf{k_1}}(\tau_2) \rho^{*}_{\mathbf{k_2}}(\tau_2) \left[\rho_{\mathbf{k_2}}(\tau_3)+ J_{\mathbf{k_2}}(\tau_3)  \right]\mathcal{G}\left(\mathbf{k_1},\tau_1-\tau_2 \right)\mathcal{G}\left(\mathbf{k_2},\tau_2-\tau_3 \right)    .   \label{ogewoon}
\end{align}
This term can be interpreted in relation to a process where an impurity creates an excitation out of the BEC at time $\tau_1$, scatters with this excitation at time $\tau_2$, and finally returns it to the BEC at time $\tau_3$. Hence we will refer to this term as the first-order scattering term. 
Every higher order term in the expansion will contain precisely one combination where every pair of functional derivatives is applied only to the exponential and contributes a power of $O_1$. A short calculation shows that these terms form the exponential power series:
\begin{equation}
     e^{-\mathcal{S}_{\textrm{eff}} / \hbar}= \left[ \sum_{n=0}^{\infty}  \frac{(-1)^n g_{ib}^n O_1^n}{n!} \right]e^{-S_{\textrm{eff}}^{F}/\hbar}.
     \label{O1series}
\end{equation}
Here, we use $O_1$ as a notation for $O_1 \left[J_{\mathbf{k}}=0 \right]$, i.e. where the source terms are set to zero. 
We can proceed to derive the second-order scattering term $O_2$. The $n=2$ term in the expansion of (\ref{pert2}) can be written as:
\begin{equation}
 \frac{1}{2!} \frac{\hbar}{n_0}     \sum_{\mathbf{k,s }}    \int_0^{\hbar \beta} d\tau  \rho_{\mathbf{k-s}}(\tau)  \frac{\delta}{\delta J_{\mathbf{k}}(\tau)}  \frac{\delta}{\delta J^*_{\mathbf{s}}(\tau)} O_1\left[J_{\mathbf{k}}\right]\left. e^{-\mathcal{S}_{\textrm{eff}}^{F} \left[J_{\mathbf{k}} \right] / \hbar}\hspace{3pt} \right|_{J_{\mathbf{k}}=0}.
 \label{pert3}
\end{equation}
Applying one of the two functional derivatives in (\ref{pert3}) to the exponential and the other to $O_1\left[J_{\mathbf{k}}\right]$ and vice versa will result in two terms that are combined in:
\begin{align}
    g_{ib}^2 O_2 \left[J_{\mathbf{k}} \right] \left. e^{-\mathcal{S}_{\textrm{eff}}^{F} \left[J_{\mathbf{k}} \right] / \hbar}\hspace{3pt} \right|_{J_{\mathbf{k}}=0},
\end{align}
where
\begin{align}
   O_2\left[J_{\mathbf{k}} \right]=&\frac{g_{ib}^2n_0}{(2\hbar V)^3 \hbar} \sum_{\mathbf{k_1,k_2,k_3}} V_{\mathbf{k_1}}^2 V_{\mathbf{k_2}}^2 V_{\mathbf{k_3}}^2 \int_{0}^{\hbar \beta} d\tau_1 \int_{0}^{\hbar \beta} d\tau_2 \int_{0}^{\hbar \beta} d\tau_3 \int_{0}^{\hbar \beta} d\tau_4 \left[  \rho^{*}_{\mathbf{k_1}}(\tau_1)+J^{*}_{\mathbf{k_1}}(\tau_1)\right] \rho_{\mathbf{k_1}}(\tau_2) \nonumber \\ & \rho^{*}_{\mathbf{k_2}}(\tau_2) \rho_{\mathbf{k_2}}(\tau_3)  \rho^{*}_{\mathbf{k_3}}(\tau_3) \left[ \rho_{\mathbf{k_3}}(\tau_4) + J_{\mathbf{k_3}}(\tau_4) \right] \mathcal{G}\left(\mathbf{k_1},\tau_1-\tau_2 \right)\mathcal{G}\left(\mathbf{k_2},\tau_2-\tau_3 \right) \mathcal{G}\left(\mathbf{k_3},\tau_3-\tau_4 \right).
\end{align}
For every term of order $n>2$ in the expansion there will be a combination of functional derivatives that will result in a contribution $\sim (n)(n-1)O_1^{n-2} O_2$. 
Performing the explicit calculation shows that these terms can be combined with the series of first-order scattering terms in (\ref{O1series}) in the following way:
\begin{equation}
     e^{-\mathcal{S}_{\textrm{eff}} / \hbar}= \left[ \sum_{n=0}^{\infty}  \frac{(-1)^n g_{ib}^n O_1^n}{n!} \right] \left( 1+g_{ib}^2 O_2\right)e^{-S_{\textrm{eff}}^{F}/\hbar}.
     \label{O2series}
\end{equation}
We will return to the factorization pattern appearing in (\ref{O2series}) later.
First we have to address the terms that are not included in this reasoning.

\subsubsection{Excitation bath terms}

In the $n=2$ order of the expansion we could have also chosen to apply both functional derivatives to $O_1\left[J_{\mathbf{k}}\right]$ in expression (\ref{pert3}), which would result in:
\begin{align}
    g_{ib}^2 \frac{\tilde{O}_1}{2}  e^{-\mathcal{S}_{\textrm{eff}}^{F}  / \hbar},
    \label{pert4}
\end{align}
where
\begin{align}
   \tilde{O}_1=\frac{1}{(2\hbar V)^2} \sum_{\mathbf{k_1,k_2}} V_{\mathbf{k_1}}^2 V_{\mathbf{k_2}}^2 \int_{0}^{\hbar \beta} d\tau_1 \int_{0}^{\hbar \beta} d\tau_2 \rho_{\mathbf{k_1}}(\tau_1) \rho^{*}_{\mathbf{k_2}}(\tau_1) \rho_{\mathbf{k_2}}(\tau_2) \rho^{*}_{\mathbf{k_1}}(\tau_2)  \mathcal{G}\left(\mathbf{k_1},\tau_1-\tau_2 \right)\mathcal{G}\left(\mathbf{k_2},\tau_1-\tau_2 \right). \label{otilde1}
\end{align}
This term can be related to a process where an impurity exchanges momentum with the excitation bath without coupling to the BEC, i.e. without creating an excitation from the condensate or scattering it back to the condensate. Whereas the scattering terms discussed in the previous subsection, such as (\ref{ogewoon}), are proportional to $n_0$ (the number of atoms in the condensate), the enhancement factor $n_0$ is absent in the excitation bath terms such as (\ref{otilde1}).

The higher-power contributions of this term will present themselves as $\tilde{O}_1^{n/2}$ in all the even $n>2$ expansion terms and together with (\ref{pert4}) they form an exponential power series as well:
\begin{equation}
     e^{-\mathcal{S}_{\textrm{eff}} / \hbar}= \left[ \sum_{n=0}^{\infty}  \frac{1}{n!} \left(g_{ib}^2 \frac{\tilde{O}_1}{2}\right)^n \right]e^{-S_{\textrm{eff}}^{F}/\hbar}.
     \label{O1series2}
\end{equation}
Just like for the scattering terms, this reasoning can be extended to higher order terms $\tilde{O}_n$ or to the combination of these terms with the scattering terms such as $\tilde{O}_1 O_1^{n-2}$. 
In general, terms uncoupled from the condensate, arise when both functional derivatives $\frac{\delta}{\delta J_{\mathbf{k}}(\tau)}  \frac{\delta}{\delta J^*_{\mathbf{s}}(\tau)}$ in a pair in  (\ref{pert2}) are applied to the two source terms contained in a scattering term $O_n\left[J_{\mathbf{k}}\right]$.
As mentioned above, the main difference between the scattering terms and the excitation bath terms is that the former contain a leading order $n_0$, while the latter do not and are therefore suppressed by a relative factor of $a_{bb} / \xi$, with $a_{bb}$ the boson-boson scattering length and $\xi$ the coherence length of the BEC.
Within the range of validity of the Bogoliubov approximation, i.e. $(n_0 a_{bb}^3) \ll 1$, they are negligible. 
Note that a similar argument has been made in a perturbative calculation \cite{ChristensenPert} to ignore diagrams where the impurity couples to bosons outside of the BEC. 
We will therefore not include these excitation bath terms in the rest of our calculations.
\subsubsection{Result}
In the discussion of the scattering terms above we have found that the power series in the first-order terms $O_1$ and all the product terms $O_2 O_1^{n-2}$ compactly factorize in expression (\ref{O2series}). An explicit calculation shows that this factorization pattern extends to higher-order scattering terms $O_n$:
\begin{equation}
e^{-\mathcal{S}_{\textrm{eff}} / \hbar } = \left( \sum_{n=0}^{\infty}  \frac{(-1)^n g_{ib}^n}{ n!} O_1^{n} \right) \times \left(  1 + g_{ib}^2 O_2 + \frac{g_{ib}^4}{2!} O_2^2 + ... \right) \times \left(  1 - g_{ib}^3 O_3 +  ... \right) \times \left( ... \right) e^{-\mathcal{S}_{\textrm{eff}}^{F} / \hbar}.
\label{pattern1}
\end{equation}
Here $O_n$ represents an $n$-th order scattering process where an impurity creates an excitation out of the BEC and scatters with it $n$ times before scattering it into the condensate again.
\begin{equation}
O_n = \frac{g_{ib}^2 n_0}{ \hbar \left(2 V \hbar  \right)^{n+1}}  \int_0^{\hbar \beta} d\tau_1 ... \int_0^{\hbar \beta} d\tau_{n+2} \sum_{\mathbf{k_1,...,k_{n+1}}} \left( \prod_{j=1}^{n+1} V_{\mathbf{k_j}}^2 \mathcal{G}_{\mathbf{k_j}} \left( \tau_{j+1}-\tau_j \right) \rho_{\mathbf{k_{j}}}(\tau_j)^* \rho_{\mathbf{k_{j}}}(\tau_{j+1})\right).
\label{On}
\end{equation}
The factorization pattern appearing in (\ref{pattern1}) suggests that the effective action can be written as:
\begin{equation}
\mathcal{S}_{\textrm{eff}} = \mathcal{S}_{\textrm{eff}}^{F} - \hbar \sum_{n=1}^{\infty}(-1)^n g_{ib}^{n} O_{n}.
\label{effaction}
\end{equation}
It might be illustrative to point out that even within the conventional Fr\"{o}hlich model a similar structure can be observed. Performing a perturbative expansion of the Fr\"{o}hlich contribution in (\ref{Lag1}) with respect to the free impurity yields:
\begin{equation}
    e^{-\mathcal{S}_{\textrm{eff}}^{F} / \hbar }=\left( 1 + O_{0} + \frac{1}{2!} O_{0}^2 + ... \right) e^{-S_{\textrm{free}} /  \hbar } ,
\end{equation}
where $S_{\textrm{free}}$ is the action functional of a non-interacting impurity and $O_0$ is a ``zeroth-order'' scattering term characterizing an impurity that creates an excitation and absorbs it a time later, without any interaction in between.
This series can be recombined in the effective action to obtain precisely the action in expression (\ref{effectiveactionFrohlich}). 

Applying the Jensen-Feynman inequality (\ref{JensenInequality}) to the effective action (\ref{effaction}) will yield a variational free energy that contains impurity density  correlation functions at different times, of an arbitrarily large order corresponding to the number of scattering events in the scattering terms. This can explicitly be seen from the product in (\ref{On}).

\subsection{Random phase approximation} 

To proceed analytically, an approximation of the impurity density correlations can be made. Relative to the model system $\mathcal{S}_0$, the impurity density correlation between the creation of an excitation at time $\tau_j$ and its absorption at time $\tau_{j+1}$ depends only on the absolute value of the time step \cite{Tempere2009Feynman}:
\begin{equation}
    \expval{\rho^*_{\mathbf{k}_{j}}(\tau_j) \rho_{\mathbf{k}_{j}}(\tau_{j+1})}_0 = \mathcal{F}_{\mathbf{k}_j} \left( |\tau_{j}-\tau_{j+1}| \right).
    \label{2correlation}
\end{equation}
$\mathcal{F}_{\mathbf{k}}(u)$ is the memory function of the impurity:
\begin{equation}
    \mathcal{F}_{\mathbf{k}}(u)= \exp \left(  \frac{\hbar k^2}{2\left( M + m_I \right)} \left[ \frac{u^2}{\hbar \beta} - u + \frac{M}{ \Omega  m_I} \frac{\cosh \left( \Omega \left[ \hbar \beta/2 -u \right]\right)-\cosh \left( \hbar \beta \Omega /2 \right)}{\sinh \left( \hbar \beta \Omega /2 \right)}\right]\right).
\end{equation}
We consider a random phase approximation (RPA) where the dominant contribution to the correlation of a number of subsequent scattering events is given by the correlations within one scattering event (\ref{2correlation}):
\begin{equation}
\expval{\prod_{j=1}^{n+1} \rho^*_{\mathbf{k}_{j}}(\tau_j) \rho_{\mathbf{k}_{j}}(t_{j+1})}_0 \approx \prod_{j=1}^{n+1} \expval{ \rho^*_{\mathbf{k}_{j}}(\tau_j) \rho_{\mathbf{k}_{j}}(\tau_{j+1})}_0 = \prod_{j=1}^{n+1} \mathcal{F}_{\mathbf{k}_j} \left( |\tau_{j}-\tau_{j+1}| \right).
\end{equation}
 Using $\mathcal{G}_{\mathbf{k}_j} \left(\hbar \beta-u \right) = \mathcal{G}_{\mathbf{k}_j} \left(u \right) $ and $\mathcal{F}_{\mathbf{k}_j} \left( \hbar \beta-u \right) = \mathcal{F}_{\mathbf{k}_j} \left(u \right) $ it is not difficult to show that within this approximation the additional contributions to the effective action constitute a power series:
\begin{equation}
 \expval{\mathcal{S}_{\textrm{eff}}}_0 = \expval{\mathcal{S}_{\textrm{eff}}^{F}}_0  + g_{ib} n_0  \hbar \beta \sum_{n=2}^{\infty}  \left( -  \frac{g_{ib}}{\hbar V}\sum_{\mathbf{k}} V_{\mathbf{k}}^2 \int_0^{\hbar \beta/2} du \mathcal{G}_{\mathbf{k}}\left( u \right) \mathcal{F}_{\mathbf{k}} \left( u \right) \right)^n .
\end{equation}
After substituting the expectation value of the Fr\"{o}hlich effective action (\ref{effectiveactionFrohlich}) with respect to the variational model system and adding the first-order energy contribution $g_{ib}n_0$ the expectation value of the full effective action becomes:
\begin{align}
     \expval{\mathcal{S}_{\textrm{eff}}}_0 = \expval{ \int_0^{\hbar \beta}\frac{m_I \dot{\mathbf{r}}^2}{2}}_0 +g_{ib} n_0  \hbar \beta \sum_{n=0}^{\infty}  \left( -  \frac{g_{ib}}{\hbar V}\sum_{\mathbf{k}} V_{\mathbf{k}}^2 \int_0^{\hbar \beta/2} du \mathcal{G}_{\mathbf{k}}\left( u \right) \mathcal{F}_{\mathbf{k}} \left( u \right) \right)^n .
     \label{action2}
\end{align}
To regularize the series we substitute the full Lippmann-Schwinger equation:
\begin{equation}
    g_{ib}^{-1}=  \frac{\mu}{2 \pi \hbar^2 a_{ib}} - \frac{1}{V} \sum_{\mathbf{k}} \frac{2 \mu}{\hbar^2 k^2},
    \end{equation}
where $\mu=\left( m_b^{-1} + m_I^{-1} \right)^{-1}$ is the reduced impurity-boson mass and $a_{ib}$ is the impurity-boson scattering length. The free energy $F_0$ of the model system and expectation value of the action of the model system $\expval{\mathcal{S}_0}_0$ can be computed. Substitution into the Jensen-Feynman inequality (\ref{JensenInequality}) yields the following variational free energy: 
\begin{align}
     F = &\frac{3}{\beta} \ln \left[ \sinh\left( \frac{\hbar \beta \Omega}{2}\right) \right] - \frac{3}{\beta} \ln \left[ \sinh\left( \frac{\hbar \beta W}{2}\right) \right]  - \frac{3}{2\beta} \ln\left( \frac{m_I+M}{m_I} \right) \nonumber \\
     &- \frac{3}{2 \beta} \frac{M}{M+m_I} \left[ \frac{\hbar \beta \Omega}{2} \coth \left( \frac{\hbar \beta \Omega}{2}\right)-1 \right] +\frac{   2\pi \hbar^2 n_0}{\mu} \frac{ 1 }{   a_{ib}^{-1} -   a_0^{-1}(M,\Omega,\beta)  }.
     \label{FreeEnergy}
\end{align}

The variational parameters are $\Omega$ and $M$ and the relation between $\Omega$ and the original oscillator frequency in the model system is given by $\Omega=W\sqrt{1+ M/m_I}$, see \cite{Feynman1955,FeynmanStatisticalMechanics,Tempere2009Feynman} for a detailed description. The free energy (\ref{FreeEnergy}) is written in a suggestive form to make the analogy with the resonance shift observed in \cite{SchadilovaDynamics,DemlerRG2017}. The resonance shift is UV convergent and in our case depends on both the temperature and the variational parameters:
\begin{equation}
    a_0^{-1}(M,\Omega,\beta) = \frac{2 \pi \hbar^2}{\mu V} \left[  \sum_{\mathbf{k}} \frac{2 \mu}{\hbar^2 k^2}- \frac{1}{\hbar }\sum_{\mathbf{k}} V_{\mathbf{k}}^2 \int_0^{\hbar \beta/2} du \mathcal{G}_{\mathbf{k}}\left( u \right) \mathcal{F}_{\mathbf{k}} \left( u \right)   \right] .
\end{equation}
The free energy (\ref{FreeEnergy}) contains the first-order energy contribution $2 \pi \hbar^2 a_{ib} n_0 / \mu$ as well, which is not included in the expression given in \cite{Tempere2009Feynman}. As a consistency check we consider the limit of weak coupling with a simplified model system where the phonon mass of the Feynman model approaches zero, $M \rightarrow 0$, while the spring constant $M W^2$ remains fixed. At zero temperature
($\beta \rightarrow \infty$), the energy in the weak coupling limit $E_{\text{weak}}$ is independent of the variational parameters and given by:
\begin{equation}
    E_{\text{weak}}= \frac{   2\pi \hbar^2 n_0}{\mu} \frac{ 1 }{   a_{ib}^{-1} -   a^{-1}_{0,\text{weak}}  }.
    \label{weakcoupling}
\end{equation}
In this limit the $u$-integral in $a^{-1}_{0,\text{weak}}$ can be analytically performed:
\begin{equation}
    a^{-1}_{0,\text{weak}} = \frac{2 \pi \hbar^2}{\mu V} \left[  \sum_{\mathbf{k}} \frac{2 \mu}{\hbar^2 k^2}- \sum_{\mathbf{k}}  \frac{V_{\mathbf{k}}^2}{ \hbar \omega_{\mathbf{k}}+ \frac{\hbar^2 k^2}{2m_I}} \right] .
    \label{weakresonanceshift}
\end{equation}
Expression (\ref{weakcoupling}) with the resonance shift (\ref{weakresonanceshift}) is precisely the mean-field result including extended interactions at zero polaron momentum $\mathbf{P}=0$ \cite{SchadilovaDynamics,DemlerRGFrohlich,VanLoon2018}. 

Feynman's path-integral formalism allows us to calculate an effective mass for the polaron $m_{\textrm{pol}}$ and a root mean square (RMS) estimate of the polaron size $\sqrt{\expval{\mathbf{r}^2}}$. 
The expression for the polaron radius depends only on the model system and remains the same as in \cite{Tempere2009Feynman}:
\begin{equation}
    \expval{r^2} =  \frac{3 \hbar }{2\Omega} \frac{m_I+M}{ m_I M} \coth \left( \frac{\hbar \beta \Omega}{2} \right). 
    \label{RMSradius}
\end{equation}
%
An expression for the effective mass can be derived by introducing a boost to the memory function of the system $\expval{ \exp\left( i \mathbf{k} \left[ \mathbf{r}(\tau) - \mathbf{r}(\sigma) \right]\right) }_0 \rightarrow \expval{ \exp\left( i \mathbf{k} \left[ \mathbf{r}(\tau) - \mathbf{r}(\sigma) \right]\right) }_0 \times \exp \left( i \mathbf{k} \cdot \mathbf{v} \left(\tau - \sigma \right)\right)$. This method is used in Feynman's seminal work on polarons within the Fr\"{o}hlich model \cite{Feynman1955,FeynmanStatisticalMechanics}. However, to incorporate the effects on the effective mass of the extended Fr\"{o}hlich contributions we only apply this boost to the effective action after the RPA contributions have been separated in (\ref{action2}). After deriving the energy as a function of $\mathbf{v}$ and expanding it up to $\mathbf{v}^2$, the factor in front of $\mathbf{v}^2 /2 $ can be identified as the polaron effective mass:
\begin{equation}
   m_{\textrm{pol}}=m_I + \lim_{\beta\rightarrow \infty} \frac{4}{3} \frac{\pi^2 \hbar^3 n_0}{\mu^2} \frac{\Gamma (M,\Omega, \beta)}{\left( a_{ib}^{-1} - a_{0}^{-1}(M,\Omega,\beta) \right)^2},
    \label{effectivemass}
\end{equation}
where $\Gamma$ is given by:
\begin{equation}
    \Gamma(M,\Omega,\beta)=\frac{1}{V} \sum_{\mathbf{k}} k^2 V_{\mathbf{k}}^2 \int_0^{\infty} du \hspace{2pt} u^2 \mathcal{G}_{\mathbf{k}}\left( u \right) \mathcal{F}_{\mathbf{k}}\left(u \right).
 \end{equation}
As far as we are aware, Feynman's prescription is only valid in the low temperature limit, so the limit $\beta \rightarrow \infty$ in expression (\ref{effectivemass}) must be taken. The effect of temperature is then estimated through the implicit temperature dependence of the variational parameters $M$ and $\Omega$, as has been done in \cite{CasteelsReducedDimensions}. Note that in the limit of $M\rightarrow 0$ in the model system, our result for the effective mass reduces to the extended Fr\"{o}hlich interactions mean-field result \cite{VanLoon2018} which suggests that corrections beyond the mean-field level are captured within this method.

\section{Results}
\label{section3}
\subsection{Comparison with the Fr\"{o}hlich model for the repulsive polaron}

\begin{figure}
\includegraphics[width=0.6\textwidth]{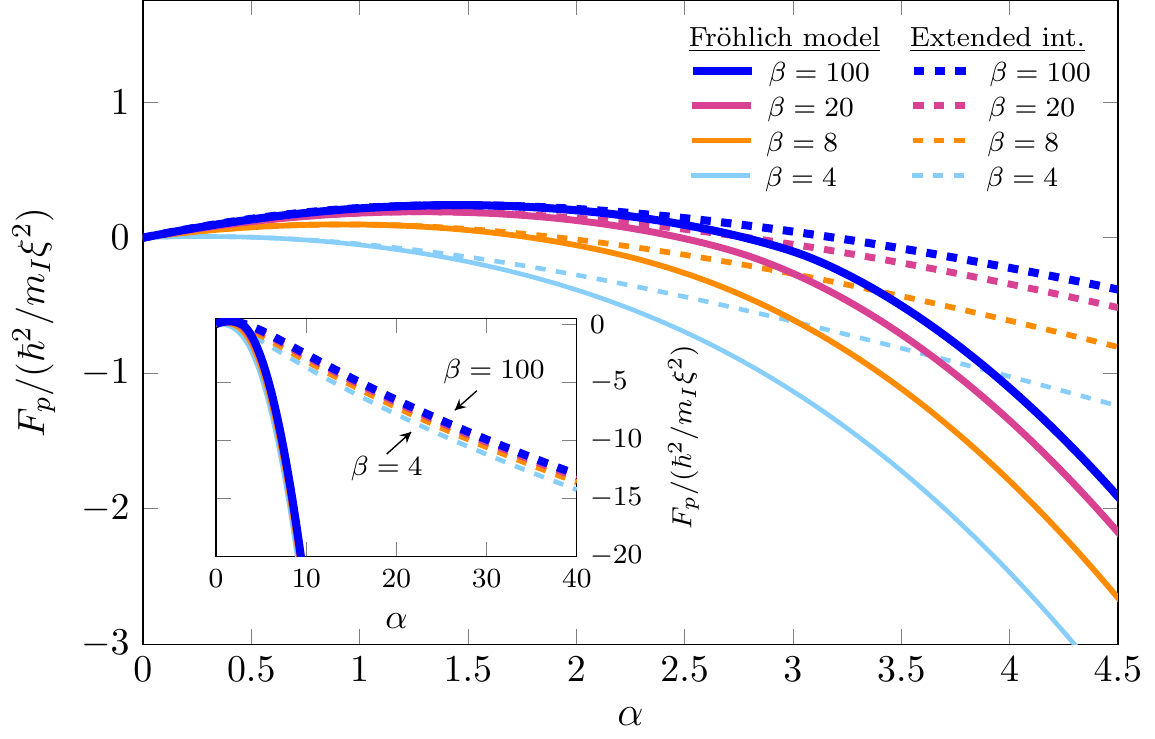}
\caption{A comparison of the polaronic contribution to the free energy including extended interactions (dashed lines) with that of the Fr\"{o}hlich model (filled lines) as a function of the coupling constant $\alpha$. The results are plotted at various temperatures $\beta=\hbar^2 /(m_I k_B T \xi^2 )=$ [4, 8, 20, 100] respectively using light blue, orange, magenta, and dark blue lines (light gray to dark gray). For the purpose of the comparison with \cite{Tempere2009Feynman}, the same impurity-condensate parameters are taken:  $m_b=3.8 \hspace{2pt} m_I$, $\xi=450 \text{ nm}$, and $ a_{bb}=2.8 \text{ nm}$. The inset shows the same results at stronger coupling.} 
\label{FreeEnergyUnits1}
\end{figure}

First, we make a direct comparison between the results obtained with Feynman's variational description within the Fr\"{o}hlich model in \cite{Tempere2009Feynman} and the results including extended interactions derived in the previous section. Because the polaronic contribution to the free energy within the Fr\"{o}hlich model is the same on both sides of the resonance we will only consider the repulsive polaron in this section. It is important to note that on this side of the resonance various shallow bound states do exist at lower energies \cite{SPRath,SchadilovaDynamics} and we are only retrieving the energy of the repulsive branch in our approach. The rich physics of Efimov bound states for an impurity in a BEC \cite{BruunEfimov,SunEfimov,YoshidaEfimov} is not expected to be captured in this approach.
Within the Fr\"{o}hlich model, the results at a given temperature can be expressed as a function of a single dimensionless polaronic coupling constant $\alpha=a_{ib}^2 / (\xi a_{bb})$. However, for (\ref{FreeEnergy}) this is no longer the case, as the results
depend also explicitly on $a_{bb}$. Nevertheless, for a fixed $a_{bb}$ we can still plot our results as a function of $\alpha$ at the repulsive side of the resonance for the purpose of the comparison. A physical cutoff corresponding to the range of the interatomic interaction is used, given by $\Lambda_c \approx 200 \xi^{-1} $ for the current system. 

\Cref{FreeEnergyUnits1} shows the results for the polaronic contribution to the free energy $F_{p}=F-2 \pi \hbar^2 a_{ib} n_0 / \mu$, in polaronic units ($\xi=m_I=\hbar=1$) at various temperatures. At weak coupling both results coincide but they start to significantly differ around $\alpha \approx 3.5 $ where the Fr\"{o}hlich model predicts a very steep decrease in energy, indicative of self-trapping. The extended interactions appear to moderate this into a much slower linear decrease of the free energy. The decrease of the polaronic contribution is even slower than the increase of the first-order contribution $2 \pi \hbar^2 a_{ib} n_0 / \mu$, and the full polaron energy for the extended interactions model never becomes negative.

\begin{figure}
\includegraphics[width=0.6\textwidth]{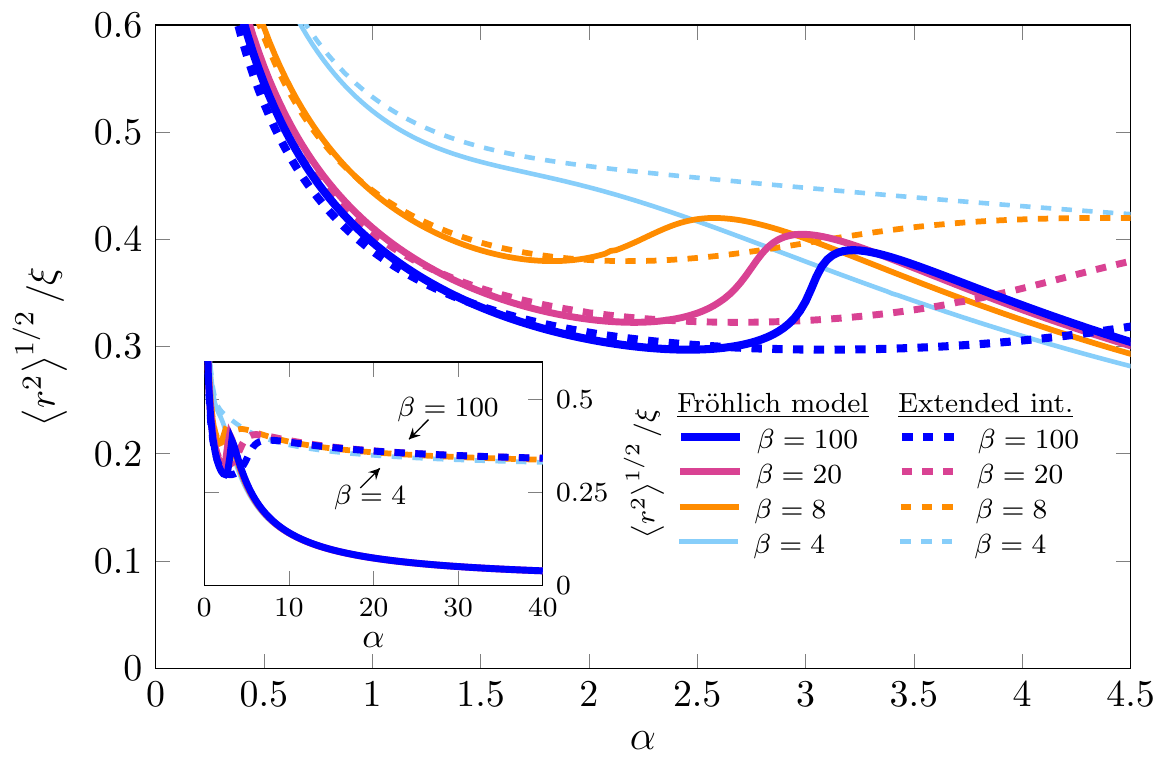}
\caption{A comparison of the RMS polaron radius of Feynman's approach including extended interactions (dashed lines) with that of the Fr\"{o}hlich model (filled lines) as a function of the coupling constant $\alpha$. The same impurity-gas parameters are used as in \cref{FreeEnergyUnits1}. The results are plotted at various temperatures $\beta=\hbar^2 /(m_I k_B T \xi^2 )=$ [4, 8, 20, 100] respectively using light blue, orange, magenta, and dark blue lines (light gray to dark gray).} 
\label{PolaronRadiusUnits1}
\end{figure}

\begin{figure}
\includegraphics[width=0.6\textwidth]{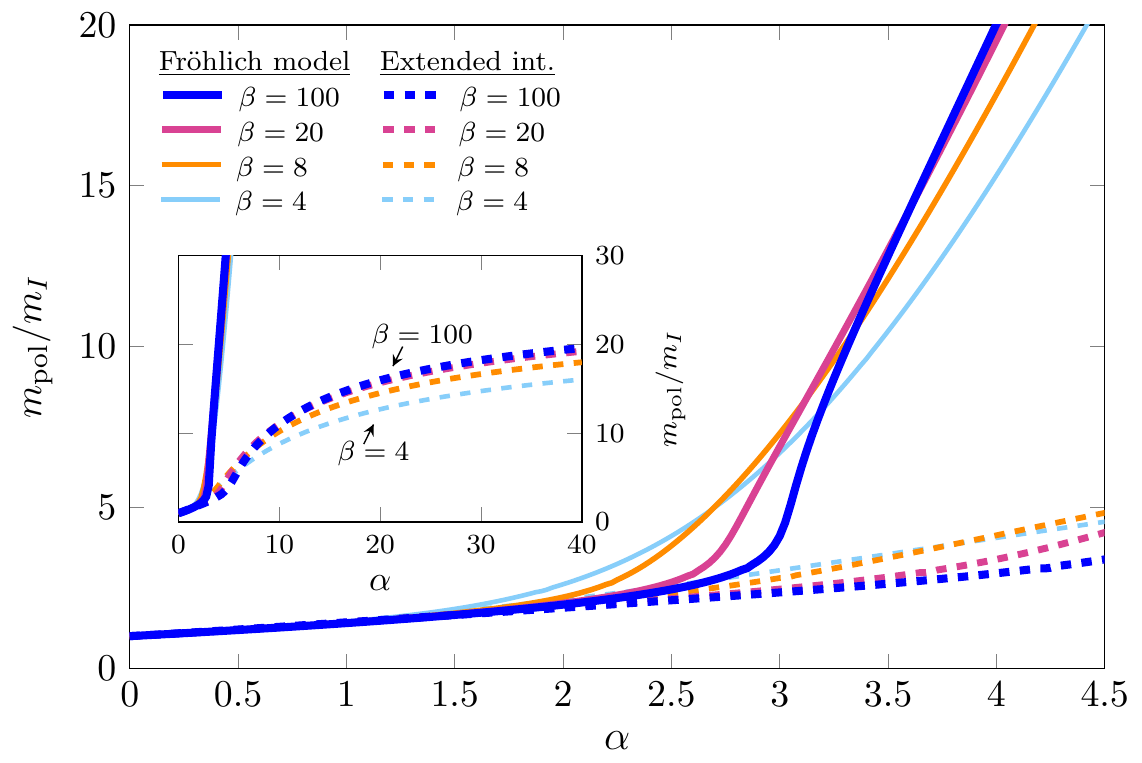}
\caption{A comparison of the polaron effective mass of Feynman's approach including extended interactions (dashed lines) with that of the Fr\"{o}hlich model (filled lines) as a function of the coupling constant $\alpha$. The same impurity-gas parameters are used as in \cref{FreeEnergyUnits1}. The results are plotted at various temperatures $\beta=\hbar^2 /(m_I k_B T \xi^2 )=$ [4, 8, 20, 100] respectively using light blue, orange, magenta, and dark blue lines (light gray to dark gray).} 
\label{PolaronMassUnits1}
\end{figure}

\Cref{PolaronRadiusUnits1} presents a comparison of the polaron RMS radius (\ref{RMSradius}) between the two models. The first noticeable difference is that the sharp kink within the Fr\"{o}hlich model, previously identified with the transition into the strong coupling regime around $\alpha=3.5$, is replaced by a smoother non-monotonic transition due to the extended interactions. Most significant is the difference at extremely strong coupling however. The inclusion of the extended interactions disproves previous predictions of the asymptotically shrinking Bose polaron within this method, which was suggestive of self-trapping as well, and shows that the polaron radius approaches a finite non-zero value around $ \approx 0.35 \xi$. Comparable conclusions follow for the effective mass of the polaron (\ref{effectivemass}), shown in \cref{PolaronMassUnits1}. The effective mass no longer exhibits a sudden and steep transition into the strong coupling regime. A period of faster increase of the effective mass is still observed around $\alpha=5 \text{-} 10$, but flattens out towards a value of roughly $\approx 20 m_I$ at even stronger coupling. Furthermore we can see that in the case of a light impurity such as considered here, the effective mass is more sensitive to temperature differences than the energy and radius. It has been pointed out that measurements of the effective mass of the polaron are expected to be particularly useful to discern between various theoretical models \cite{DemlerRGFrohlich,GrusdtReview}. Based on our results we expect this to be even more the case when the temperature dependence is measured as well.

The converging effective mass and polaron radius together with the positive free energy suggest that self-trapping does not take place for the repulsive polaron when the extended interactions are included. This is qualitatively in agreement with the findings of the RG approach \cite{DemlerRG2017}, where no self-trapping is observed for the repulsive polaron. 

\subsection{Comparison with other theoretical results}

In this subsection we provide a comparison with other recent results in the literature, specifically with the mean-field approach \cite{SchadilovaDynamics}, the RG approach \cite{DemlerRG2017} and Quantum Monte Carlo calculations \cite{ArdilaMC2019}. As mentioned above, the mean-field expressions for the energy and effective mass \cite{SchadilovaDynamics,VanLoon2018} can also be obtained from the weak coupling limit of Feynman's model. 

Before proceeding to the discussion, one aspect of the Feynman model has to be addressed. As can be seen from expression (\ref{FreeEnergy}), the variational landscape can contain poles where the free energy diverges to negative (or positive) infinity, accompanied by a divergence of the effective mass (\ref{effectivemass}). On the negative side of the resonance, even at weaker coupling, these poles are present. However, below a critical coupling strength there exists a separated local minimum that corresponds to the polaronic state. To plot the polaron energy of the attractive branch we follow this local minimum starting from weak interactions up to the point where it merges with one of the aforementioned poles, at which both the energy and effective mass diverge. 

To better understand the physical significance of these divergences, it is illustrative to observe that the same type of pole is present in the extended mean-field treatment \cite{SchadilovaDynamics,VanLoon2018}, where it is independent of any additional variational parameters. In the MF model this divergence can be shown to be accompanied by a rapid depletion of the BEC, which is no longer accurately described within Bogoliubov theory. We therefore believe that the poles observed within our treatment can be interpreted as a runaway pathway related to the shortcomings of the Bogoliubov approximation. A detailed discussion of similar divergences, observed in RG theory, is presented in \cite{DemlerRG2017}. Note that in \cite{YoshidaEfimov} no divergences are observed for the polaron at unitarity, which indicates that a correct treatment of Efimov physics is of importance here as well. 

For the repulsive polaron no runaway pathways exist at weak coupling and we simply follow the global minimum of the variational landscape. Only at extremely strong coupling, separated divergences start to appear and the polaronic state becomes a local minimum. This local minimum continues to exist across the resonance towards negative scattering lengths, and it is not clear if we can interpret it as the repulsive polaron state from this point on. For the purpose of the comparison with RG in this subsection, we will restrict our study of the repulsive branch to couplings below the critical coupling presented in \cite{DemlerRG2017}. At this point the aforementioned transition into a local minimum has not yet taken place.

\begin{figure}
\includegraphics[width=0.6\textwidth]{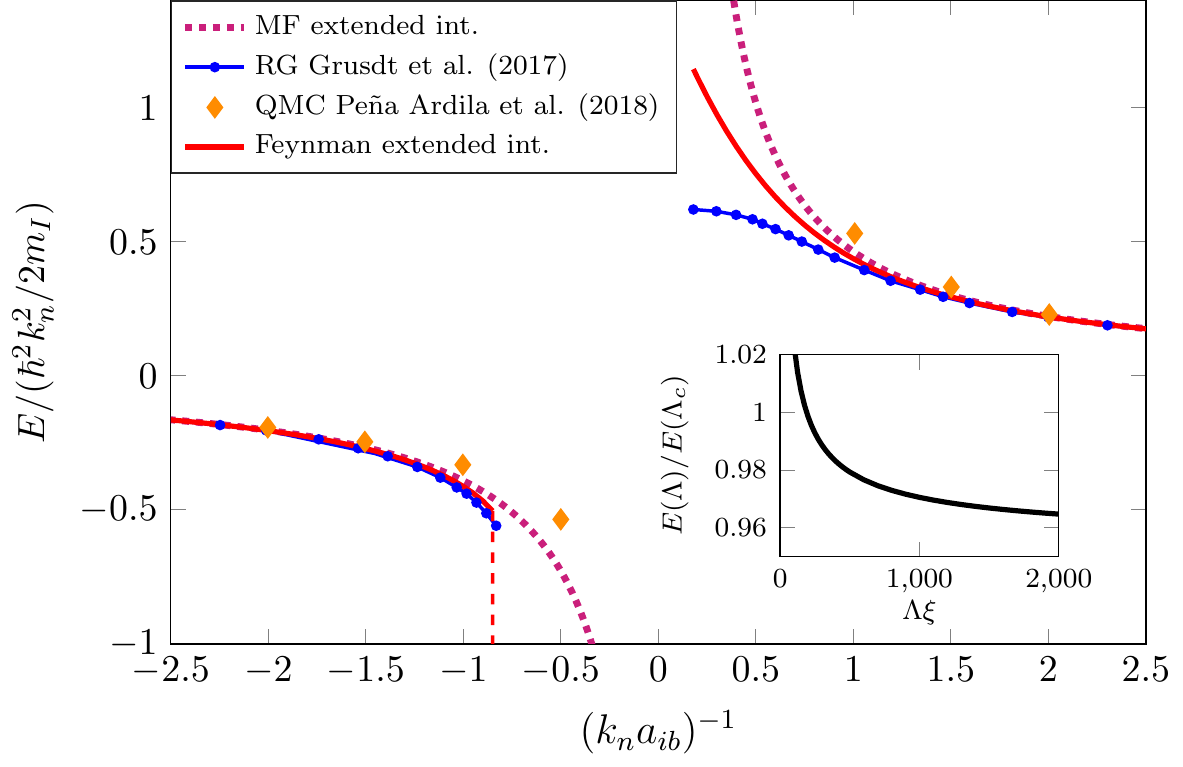}
\caption{A comparison of the polaron energy obtained with the path-integral variational method including extended interactions (solid), the mean-field model including extended interactions \cite{SchadilovaDynamics} (dotted), the RG approach \cite{DemlerRG2017} (connected dots) and QMC \cite{ArdilaMC2019} (diamonds). The impurity and condensate parameters correspond to the experiment of J{\o}rgensen et al. \cite{Jorgensen}, given by $m_I=m_b$ and $a_{bb}=9 a_0$, $a_0$ being the Bohr radius. We take a UV momentum cutoff of the range of the Feshbach resonance $\Lambda_c=(60 a_0)^{-1}\approx 190 \xi^{-1}$ in this experiment, given in \cite{Jorgensen} and also used in \cite{DemlerRG2017}. On the figure the inverse scattering length is measured in terms of $k_n=\left(6\pi^2 n_0 \right)^{1/3}$. The temperature integral cutoff corresponds to $\beta_c=\hbar^2 /(\xi^2 m_I k_B T )=200$ or $0.17$ nK. For the purpose of the comparison with RG we plot the repulsive branch up to $(k_n a_{ib})^{-1} \approx 0.18$. On the attractive branch we can only show the RG data up to the lower range of fig.~9 in \cite{DemlerRG2017}. The inset shows the high cutoff behavior of $E(\Lambda) / E(\Lambda_c)$ in the Feynman approach at strong coupling for $(k_n a_{ib})^{-1}=0.3$.}  
\label{EnergyUnits2}
\end{figure}

\begin{figure}
\includegraphics[width=0.6\textwidth]{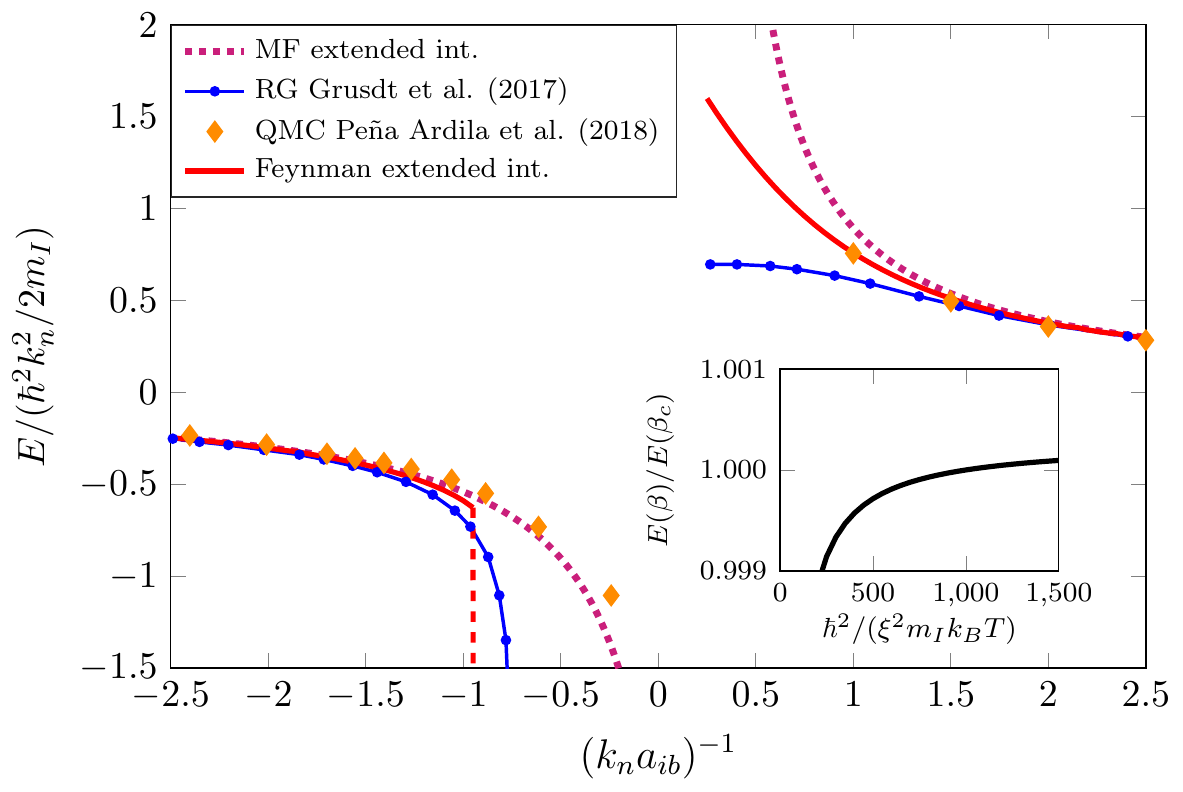}
\caption{A comparison of the polaron energy obtained with the path-integral variational method including extended interactions (solid), the mean-field model including extended interactions \cite{SchadilovaDynamics} (dotted), the RG approach \cite{DemlerRG2017} (connected dots) and QMC \cite{ArdilaMC2019} (diamonds). The impurity and condensate parameters correspond to the experiment of Hu et al. \cite{Hu}, given by $m_b=2.17m_I$ and $a_{bb}=100 a_0$, $a_0$ being the Bohr radius. We take the same UV cutoff $\Lambda_c=10^3 / \xi$ as used in \cite{DemlerRG2017}. The temperature integral cutoff corresponds to $\beta_c=\hbar^2/ (\xi^2 m_I k_B T )=1000$ or $0.3$ nK. For the purpose of the comparison with RG we plot the repulsive branch up to $(k_n a_{ib})^{-1} \approx 0.25$. The inset shows the low temperature convergence of $E(\beta)/E(\beta_c)$ in the Feynman approach at strong coupling for $(k_n a_{ib})^{-1}=0.3$. }  
\label{EnergyUnits3}
\end{figure}

\begin{figure}
\includegraphics[width=0.6\textwidth]{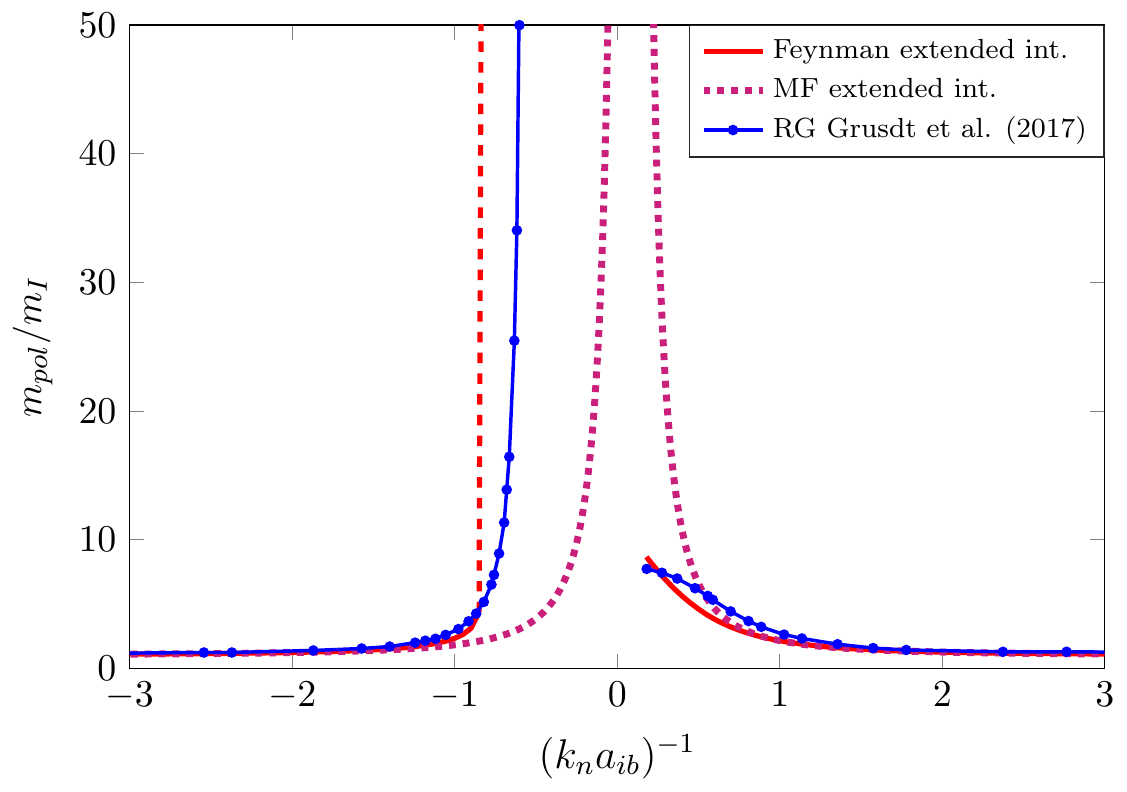}
\caption{A comparison of the polaron effective mass between Feynman's model including extended interactions (solid), the mean-field model including extended interactions \cite{SchadilovaDynamics,VanLoon2018} (dotted) and the RG approach \cite{DemlerRG2017} (connected dots). The same impurity-condensate parameters were used as in \cref{EnergyUnits2} corresponding to the experiment of J{\o}rgensen et al. \cite{Jorgensen}. For the purpose of the comparison with RG we plot the repulsive branch up to $(k_n a_{ib})^{-1} \approx 0.18$.} 
\label{PolaronMassUnits2}
\end{figure}

Figures \ref{EnergyUnits2} and \ref{EnergyUnits3} show a comparison of the polaron energies obtained with various methods, across the resonance for impurity-condensate parameters used in the experiments \cite{Jorgensen} and \cite{Hu}, respectively. To provide an accurate comparison with RG, the same respective finite values of the momentum cutoff were used as in \cite{DemlerRG2017}. As shown on the inset of \cref{EnergyUnits2}, further convergence of a few percent is expected at infinite cutoff in this case. The cutoff used for \cref{EnergyUnits3} is larger and the results are much closer to convergence. The cutoff dependence of the Feynman approach within the Fr\"{o}hlich BEC model has been discussed in \cite{Tempere2009Feynman} and \cite{Vlietinck2015}. For both calculations finite temperatures, at which the energy has converged beyond any noticeable difference in the figures, were used to represent zero temperature. This convergence is shown on the inset of \cref{EnergyUnits3}. 

For the repulsive polaron we observe a relatively good quantitative agreement with QMC data in \cref{EnergyUnits2} and an excellent agreement in \cref{EnergyUnits3}. Note, however, that the QMC calculation does not rely on the Bogoliubov approximation. Our results predict no divergence of the repulsive branch energy in contrast to the mean-field treatment, but towards stronger coupling a quantitative discrepancy with the RG approach appears. However, as shown in \cref{PolaronMassUnits2}, a much better agreement exists for the effective mass of the repulsive polaron between the two methods. One possible explanation for the discrepancy in energy is the previously discussed logarithmic divergence captured in the RG theory. The QMC study \cite{ArdilaMC2019} does not elaborate on the cutoff dependence so the status of the logarithmic divergences in this method is unclear. On the negative side of the resonance we see qualitative agreement with RG where the polaron energy and effective mass diverge at a weaker interaction strength than predicted by the mean-field description or QMC. 

The theoretical results can also be compared to the experimental data points from \cite{Hu,Jorgensen}, which we have not explicitly added to the figures for the purpose of clarity. At weak coupling all theoretical approaches are in excellent agreement with experiments. For the repulsive branch as the coupling gets stronger, the data points of J{\o}rgensen et al. \cite{Jorgensen} lie at higher energies than QMC, even after the non-homogeneity of the three-body decay processes is taken into account in the spectroscopic signal \cite{ArdilaMC2019}. As both RG and our results lie below QMC it follows that the agreement with this experiment is not close in this regime. Several reasons for this discrepancy are suggested in \cite{ArdilaMC2019}. In contrast we find excellent agreement with the experiment of Hu et al. \cite{Hu} on the repulsive branch, which is also in much closer agreement with QMC than the J{\o}rgensen et al. experiment. The early divergence for the attractive polaron that is found in our results and the RG approach is observed in neither experiment, and as mentioned above, its understanding requires a further study of the validity of the Bogoliubov approximation and the effects of Efimov physics in that regime.

\section{Conclusion}

We have studied the ground-state properties of the Bose polaron beyond the Fr\"{o}hlich paradigm using Feynman's variational path-integral formalism. For this purpose we derived the Lagrangian of an impurity immersed in a condensate within the Bogoliubov approximation. 
The extended Fr\"{o}hlich interactions take the form of quadratic position- and velocity-dependent terms in the phonon variables. 
By expanding the position-dependent terms as a full perturbative series the path integral over the phonon variables can be performed to obtain an effective action. 
This is done within the Bogoliubov approximation, neglecting perturbative terms that contain no coupling to the condensate. We do not expect the velocity-dependent terms to contribute significantly to the ground-state properties based on other theoretical studies in the literature. 
The Jensen-Feynman inequality provides a variational expression for the upper bound on the free energy. Due to the extended interactions it contains a series of impurity density correlations that, as far as we know, does not reduce to an analytic expression. 
To proceed analytically a random phase approximation is made that decomposes the higher order impurity-excitation scattering correlations as a product of subsequent scattering correlations. 
The RPA yields simple variational expressions for the polaron energy and effective mass that reduce to the extended Fr\"{o}hlich mean-field results in the weak coupling limit. 
For the repulsive polaron we compared the predictions with those of the Fr\"{o}hlich model and found that the sharp transition to the strong coupling regime, which was interpreted as a possible shortcoming of the path-integral approach for the Fr\"{o}hlich model, is now replaced by a smooth crossover suggestive of the absence of self-trapping. For the attractive polaron we observed an abrupt divergence of the energy and effective mass at a certain critical coupling strength. This is related to the local polaronic minimum getting absorbed by a runaway pathway in the variational landscape, and is interpreted as a breakdown of the Bogoliubov approximation within our approach.

Various future perspectives for this method exist. While Feynman's method in theory captures the full effect of the excitations at the level of the effective action, it relies on a simple two-parameter model system to capture their influence on the impurity at the level of the free energy and effective mass. Moreover, we invoked an additional approximation by using the random phase approximation. One future perspective would be to consider different model systems with more degrees of freedom, as has been already proposed in \cite{DemlerRGFrohlich,GrusdtReview}. 
It would also be interesting to extend our study to an impurity in a one-dimensional BEC. In \cite{GrusdtFeynman1DComparison} it has been shown that in this regime Feynman's method and the RG approach are in excellent agreement within the Fr\"{o}hlich model, and the observed discrepancies in three dimensions are not present.
Finally, at the impurity densities created in current experiments, many-polaron effects are expected to be non-negligible already on a mean-field level \cite{VanLoon2018}. The variational path-integral approach has been used to study these effects for Fr\"{o}hlich polarons in solids \cite{BipolaronVerbist,Klimin2004} and was applied to the study of the Bose bipolaron within the Fr\"{o}hlich model \cite{CasteelsBipolaron}. Combining the inclusion of the extended interactions and the approximations made in this work with these methods would open a possible avenue towards the study of bipolarons and many-polaron effects in Bose gases beyond the mean-field level with the path-integral formalism. 

\begin{acknowledgments}
T. Ichmoukhamedov acknowledges financial support in the form of a Ph.D. fellowship of the Research Foundation - Flanders (FWO), project 1135519N. This research was supported by the University Research Fund (BOF) of the University
of Antwerp and by the Flemish Research Foundation (FWO-Vl), projects G.0429.15.N and GOG66.16.N. 
\end{acknowledgments}
\appendix
\section{Derivation of the classical Lagrangian  \label{Appendix A} }

Starting from expression (\ref{Ham2}), relative to $E_0$ and the first-order energy shift, the Hamiltonian for a single impurity in the BEC can be written as:

\begin{align}
\hat{H} = & \frac{ \hat{\mathbf{p}}^2}{2m_I} +  \sum_{\mathbf{k \neq 0}} \epsilon({\mathbf{k}}) \hat{\alpha}^{\dagger}_{\mathbf{k}} \hat{\alpha}_{\mathbf{k}}+ \frac{\sqrt{N_0}g_{ib}}{V} \sum_{\mathbf{k  }}  \hat{\rho}_{\mathbf{k}} V_{\mathbf{k}} \left( \hat{\alpha}^{\dagger}_{\mathbf{-k}} + \hat{\alpha}_{\mathbf{k}} \right)  \nonumber \\
&+\frac{g_{ib}}{V}   \sum_{\mathbf{k,s }}   \hat{\rho}_{\mathbf{k-s}} W^{(1)}_{\mathbf{k},\mathbf{s}} \hat{\alpha}^{\dagger}_{\mathbf{s}} \hat{\alpha}_{\mathbf{k}} + \frac{1}{2} \frac{g_{ib}}{V}   \sum_{\mathbf{k,s }}  \hat{\rho}_{\mathbf{k-s}} W^{(2)}_{\mathbf{k},\mathbf{s}} \left( \hat{\alpha}^{\dagger}_{\mathbf{s}}  \hat{\alpha}^{\dagger}_{\mathbf{-k}}  + \hat{\alpha}_{\mathbf{k}} \hat{\alpha}_\mathbf{-s} \right). 
\label{AppendixHam2}
\end{align}
The next step is to introduce position and momentum operators defined by:
\begin{align}
&\hat{Q}_{\mathbf{k}}= \sqrt{\frac{\hbar}{2M \omega(\mathbf{k})}}\left(\hat{\alpha}_\mathbf{k}+\hat{\alpha}^{\dagger}_{\mathbf{-k}}\right), \\
&\hat{P}_{\mathbf{k}}= i \sqrt{\frac{\hbar M \omega(\mathbf{k}) }{2}}\left(\hat{\alpha}^{\dagger}_{\mathbf{k}} - \hat{\alpha}_{\mathbf{-k}} \right),
 \end{align}
 which obey $\left[ \hat{Q}_{\mathbf{k}}, \hat{P}_{\mathbf{k'}} \right] = i \hbar \delta_{\mathbf{k,k'}}$ and $\hat{Q}^{\dagger}_{\mathbf{k}}=\hat{Q}_{\mathbf{-k}}$, $\hat{P}^{\dagger}_{\mathbf{k}}=\hat{P}_{\mathbf{-k}}$. Here, $M$ is an arbitrary phonon mass and the frequency $\hbar \omega_{\mathbf{k}}=\epsilon(\mathbf{k})$ corresponds to the Bogoliubov energy dispersion. The Hamiltonian in terms of $\hat{Q}_{\mathbf{k}}$ and $\hat{P}_{\mathbf{k}}$ becomes: 
\begin{align}
\hat{H} = &  \frac{ \hat{\mathbf{p}}^2}{2m_I} +  \sum_{\mathbf{k}} \frac{M \omega_{\mathbf{k}}^2}{2} \hat{Q}^{\dagger}_{\mathbf{k}} \hat{Q}_{\mathbf{k}} + \sum_{\mathbf{k}} \frac{1}{2M} \hat{P}^{\dagger}_{\mathbf{k}} \hat{P}_{\mathbf{k}} + \frac{\sqrt{N_0}g_{ib}}{V} \sum_{\mathbf{k \neq 0 }}  \hat{\rho}_{\mathbf{k}}  \sqrt{\frac{2M \omega_{\mathbf{k}}}{\hbar}}V_{\mathbf{k}} \hat{Q}_{\mathbf{k}}  \nonumber \\
&+\frac{g_{ib}}{V}   \sum_{\mathbf{k,s }}  \hat{\rho}_{\mathbf{k-s}} V_{\mathbf{k}} V_{\mathbf{s}}  \frac{M \sqrt{ \omega_{\mathbf{k}} \omega_{\mathbf{s}} }}{2\hbar} \hat{Q}^{\dagger}_{\mathbf{s}} \hat{Q}_{\mathbf{k}} +\frac{g_{ib}}{V}   \sum_{\mathbf{k,s }}  \hat{\rho}_{\mathbf{k-s}} V_{\mathbf{k}}^{-1} V_{\mathbf{s}}^{-1} \frac{1}{2 M \hbar \sqrt{ \omega_{\mathbf{k}} \omega_{\mathbf{s}} } } \hat{P}^{\dagger}_{\mathbf{k}} \hat{P}_{\mathbf{s}}.
\label{AppendixHam3}
\end{align}

Two types of diverging terms containing the commutator $\left[\hat{Q}_{\mathbf{k}} , \hat{P}_{\mathbf{k}} \right]$ arise in the derivation of (\ref{AppendixHam3}). The first one corresponds to the ground-state energy of the introduced harmonic oscillators $- \sum_{\mathbf{k}} \hbar \omega_{\mathbf{k}} / 2 $.  The other one arises from the cross terms in the extended interactions and is given by $-\frac{g_{ib}}{2V} \sum_{\mathbf{k}} W_{\mathbf{k,k}}^{(1)} $. This term contains a UV divergence that can not be regularized by taking the cutoff-dependence of $g_{ib}$ into account. Neither of these terms contains the impurity coordinate and we will not include them in further discussion. The classical Hamiltonian corresponding to (\ref{AppendixHam3}) is obtained by replacing the operators with complex scalar variables that obey $Q_{\mathbf{k}}^{*}=Q_{\mathbf{-k}}$:
\begin{align}
H = & \frac{ \mathbf{p}^2}{2m_I} +  \sum_{\mathbf{k}} \frac{M \omega_{\mathbf{k}}^2}{2} Q_{\mathbf{-k}} Q_{\mathbf{k}} + \sum_{\mathbf{k}} \frac{1}{2M} P_{\mathbf{-k}} P_{\mathbf{k}} + \frac{\sqrt{N_0}g_{ib}}{V} \sum_{\mathbf{k \neq 0 }}  \rho_{\mathbf{k}}  \sqrt{\frac{2M \omega_{\mathbf{k}}}{\hbar}}V_{\mathbf{k}} Q_{\mathbf{k}}  \nonumber \\
&+\frac{g_{ib}}{V}   \sum_{\mathbf{k,s }}  \rho_{\mathbf{k-s}} V_{\mathbf{k}} V_{\mathbf{s}}  \frac{M \sqrt{ \omega_{\mathbf{k}} \omega_{\mathbf{s}} }}{2\hbar} Q_{\mathbf{-s}} Q_{\mathbf{k}} +\frac{g_{ib}}{V}   \sum_{\mathbf{k,s }}   \rho_{\mathbf{k-s}} V_{\mathbf{k}}^{-1} V_{\mathbf{s}}^{-1} \frac{1}{2 M \hbar \sqrt{ \omega_{\mathbf{k}} \omega_{\mathbf{s}} } } P_{\mathbf{-k}} P_{\mathbf{s}}.  
\label{AppendixHam4}
\end{align}
The Legendre transformation:
\begin{equation}
L =  \sum_{\mathbf{q}} \frac{\partial H}{\partial P_{\mathbf{q}}} P_{\mathbf{q}}  +  \frac{\partial H}{\partial \mathbf{p}} \cdot \mathbf{p} - H
\end{equation}
results in the classical Lagrangian:
\begin{align}
L = & \frac{ m_I \mathbf{\dot{r}}^2}{2} - \sum_{\mathbf{k}} \frac{M \omega_{\mathbf{k}}^2}{2} Q_{\mathbf{-k}} Q_{\mathbf{k}} + \sum_{\mathbf{k}} \frac{1}{2M} P_{\mathbf{-k}} P_{\mathbf{k}} -\frac{\sqrt{N_0}g_{ib}}{V} \sum_{\mathbf{k \neq 0 }}  \rho_{\mathbf{k}}  \sqrt{\frac{2M \omega_{\mathbf{k}}}{\hbar}}V_{\mathbf{k}} Q_{\mathbf{k}}  \nonumber \\
&-\frac{g_{ib}}{V}   \sum_{\mathbf{k,s }}  \rho_{\mathbf{k-s}} V_{\mathbf{k}} V_{\mathbf{s}}  \frac{M \sqrt{ \omega_{\mathbf{k}} \omega_{\mathbf{s}} }}{2\hbar} Q_{\mathbf{-s}} Q_{\mathbf{k}} +\frac{g_{ib}}{V}   \sum_{\mathbf{k,s }}  \rho_{\mathbf{k-s}} V_{\mathbf{k}}^{-1} V_{\mathbf{s}}^{-1} \frac{1}{2 M \hbar \sqrt{ \omega_{\mathbf{k}} \omega_{\mathbf{s}} } } P_{\mathbf{-k}} P_{\mathbf{s}}, 
\label{AppendixLag1}
\end{align}
where the impurity coordinate $\mathbf{r}$ has been introduced. The Lagrangian (\ref{AppendixLag1}) still has to be written as a function of:
\begin{equation}
\dot{Q}_{\mathbf{q}} = \frac{\partial H}{\partial P_{\mathbf{q}}} \implies  P_{\mathbf{-k}} = M \dot{Q}_{\mathbf{k}} -\frac{g_{ib}}{V}   \sum_{ \mathbf{q } } \rho_{\mathbf{q-k}} V_{\mathbf{q}}^{-1} V_{\mathbf{k}}^{-1} \frac{1}{ \hbar \sqrt{ \omega_{\mathbf{q}} \omega_{\mathbf{k}} } } P_{\mathbf{-q}}.  
\label{legendre1}
\end{equation}

To simplify the algebra we can multiply the RHS of (\ref{legendre1}) by $P_{\mathbf{k}} / (2M)$ and perform the summation over $\mathbf{k}$. The two momentum-dependent terms in the Lagrangian can then be compactly written as:
\begin{align}
L = & \frac{ m_I \mathbf{\dot{r}}^2}{2} - \sum_{\mathbf{k}} \frac{M \omega_{\mathbf{k}}^2}{2} Q_{\mathbf{-k}} Q_{\mathbf{k}} + \sum_{\mathbf{k}} \frac{1}{2} \dot{Q}_{\mathbf{k}} P_{\mathbf{k}} -\frac{\sqrt{N_0}g_{ib}}{V} \sum_{\mathbf{k \neq 0 }}  \rho_{\mathbf{k}}  \sqrt{\frac{2M \omega_{\mathbf{k}}}{\hbar}}V_{\mathbf{k}} Q_{\mathbf{k}}  \nonumber \\
&-\frac{g_{ib}}{V}   \sum_{\mathbf{k,s }}  \rho_{\mathbf{k-s}} V_{\mathbf{k}} V_{\mathbf{s}}  \frac{M \sqrt{ \omega_{\mathbf{k}} \omega_{\mathbf{s}} }}{2\hbar} Q_{\mathbf{-s}} Q_{\mathbf{k}} .  
\label{AppendixLag2}
\end{align}
Next we shall look for an explicit expression for $P_{\mathbf{k}}$. Expression (\ref{legendre1}) can be equivalently written as:
\begin{equation}
P_{\mathbf{k}} = M \dot{Q}_{\mathbf{-k}} -\frac{g_{ib}}{V}   \sum_{ \mathbf{q } } \rho_{\mathbf{k-q}} V_{\mathbf{q}}^{-1} V_{\mathbf{k}}^{-1} \frac{1}{ \hbar \sqrt{ \omega_{\mathbf{q}} \omega_{\mathbf{k}} } } P_{\mathbf{q}}.
\label{momentum1}
\end{equation}
Note that for a single impurity $\rho_{\mathbf{k-q}}=\rho_{\mathbf{k}} \rho_{\mathbf{-q}}$. By multiplying (\ref{momentum1}) with $\frac{V_{\mathbf{k}}^{-1} \rho_{\mathbf{-k}}}{\sqrt{\hbar \omega_{\mathbf{k}}}}$ and performing the summation over $\mathbf{k}$ we find:
\begin{equation}
\sum_{\mathbf{k}} \frac{V_{\mathbf{k}}^{-1} \rho_{\mathbf{-k}}}{\sqrt{\hbar \omega_{\mathbf{k}}}} P_{\mathbf{k}} = \sum_{\mathbf{k}} \frac{V_{\mathbf{k}}^{-1} \rho_{\mathbf{-k}}}{\sqrt{\hbar \omega_{\mathbf{k}}}} M \dot{Q}_{\mathbf{-k}} - \frac{g_{ib}}{V} \sum_{\mathbf{k}} \frac{V_{\mathbf{k}}^{-2}}{\hbar \omega_{\mathbf{k}}} \sum_{\mathbf{q}} \frac{V_{\mathbf{q}}^{-1} \rho_{\mathbf{-q}}}{\sqrt{\hbar \omega_{\mathbf{q}}}} P_{\mathbf{q}} .
\label{momentum2}
\end{equation}
Equation (\ref{momentum2}) can be algebraically solved to obtain:
\begin{equation}
\sum_{\mathbf{k}} \frac{V_{\mathbf{k}}^{-1} \rho_{\mathbf{-k}}}{\sqrt{\hbar \omega_{\mathbf{k}}}} P_{\mathbf{k}} =   \eta \sum_{\mathbf{k}} \frac{V_{\mathbf{k}}^{-1} \rho_{\mathbf{-k}}}{\sqrt{\hbar \omega_{\mathbf{k}}}} M \dot{Q}_{\mathbf{-k}}  ,
\label{momentum3}
\end{equation}
where $\eta=\left(1 +  \frac{g_{ib}}{V} \sum_{\mathbf{k}} \frac{V_{\mathbf{k}}^{-2}}{\hbar \omega_{\mathbf{k}}} \right)^{-1}$.  After substituting (\ref{momentum3}) into (\ref{momentum1}) the expression for $P_{\mathbf{k}}$ becomes:
\begin{equation}
P_{\mathbf{k}}=M \dot{Q}_{\mathbf{-k}} - \frac{g_{ib}}{V} M \eta \frac{V_{\mathbf{k}}^{-1} \rho_{\mathbf{k}}}{\sqrt{\hbar \omega_{\mathbf{q}}}} \sum_{\mathbf{q}} \frac{V_{\mathbf{q}}^{-1} \rho_{\mathbf{-q}}}{\sqrt{\hbar \omega_{\mathbf{q}}}}  \dot{Q}_{\mathbf{-q}} .
\label{momentum4}
\end{equation}
Finally we can substitute (\ref{momentum4}) in the Lagrangian:
\begin{align}
L = & \frac{ m_I \mathbf{\dot{r}}^2}{2} +  \frac{M}{2} \sum_{\mathbf{k}} \dot{Q}_{\mathbf{k}} \dot{Q}_{\mathbf{-k}} - \sum_{\mathbf{k}} \frac{M \omega_{\mathbf{k}}^2}{2} Q_{\mathbf{-k}} Q_{\mathbf{k}}  -\frac{\sqrt{N_0}g_{ib}}{V} \sum_{\mathbf{k \neq 0 }}  \rho_{\mathbf{k}}  \sqrt{\frac{2M \omega_{\mathbf{k}}}{\hbar}}V_{\mathbf{k}} Q_{\mathbf{k}}  \nonumber \\
&-\frac{g_{ib}}{V} \frac{M}{2}  \sum_{\mathbf{k,s }}  \rho_{\mathbf{k-s}} V_{\mathbf{k}} V_{\mathbf{s}}  \frac{ \sqrt{ \omega_{\mathbf{k}} \omega_{\mathbf{s}} }}{\hbar} Q_{\mathbf{-s}} Q_{\mathbf{k}} -\frac{g_{ib}}{V} \frac{M \eta}{2} \sum_{\mathbf{k,s}} \frac{V_{\mathbf{k}}^{-1} V_{\mathbf{s}}^{-1}}{\hbar \sqrt{\omega_{\mathbf{k}} \omega_{\mathbf{s}}}}  \rho_{\mathbf{k}-\mathbf{s}}\dot{Q}_{\mathbf{k}}\dot{Q}_{\mathbf{-s}}.  
\label{AppendixLag3}
\end{align}

\bibliography{refs}

\end{document}